\documentclass[aps,superscriptaddress,floats,showpacs,floatfix]{revtex4}
\usepackage{bm}
\usepackage{graphicx}
\usepackage{subfigure}
\usepackage{amssymb}
\usepackage{amsmath}
\usepackage{color}
\usepackage[a4paper,left=1.5cm, right=1.3cm, top=2cm,bottom=2cm]{geometry}

\begin{document}

\title{Predicting transition ranges to turbulent viscous boundary layers \\in low Prandtl number convection flows}
\author{Janet D. Scheel}
\affiliation{Department of Physics, Occidental College, 1600 Campus Road, M21, Los Angeles, California 90041, USA}
\author{J\"org Schumacher}
\affiliation{Institut f\"ur Thermo- und Fluiddynamik, Technische Universit\"at Ilmenau, Postfach 100565, D-98684 Ilmenau, Germany}
\date{\today}

\begin{abstract}
We discuss two aspects of turbulent Rayleigh-B\'{e}nard convection (RBC) on the basis of high-resolution direct numerical 
simulations in a unique setting; a closed cylindrical cell of aspect ratio of one. First, we present a comprehensive comparison of statistical 
quantities such as energy dissipation rates and boundary layer thickness scales. Data are used from three simulation run series at Prandtl 
numbers $Pr$ that cover two orders of magnitude. In contrast to most previous studies in RBC the focus of the present work is on convective 
turbulence at very low Prandtl numbers including $Pr=0.021$ for liquid mercury or gallium  and $Pr=0.005$ for liquid sodium. In this parameter range of 
RBC, inertial effects cause a dominating turbulent momentum transport that is in line with highly intermittent fluid turbulence both in the bulk 
and in the boundary layers and thus should be able to trigger a transition to the fully turbulent boundary layers of the ultimate regime of convection for higher Rayleigh number. Secondly, we predict the  ranges of  Rayleigh numbers for which the viscous boundary layer will transition to turbulence 
and the flow as a whole will cross over into the ultimate regime. These transition ranges are obtained by extrapolation from our simulation data. 
The extrapolation methods are based on the large-scale properties of the velocity profile. Two of the three methods predict similar ranges for 
the transition to ultimate convection when their uncertainties are taken into account. All three extrapolation methods indicate that the range of 
critical Rayleigh numbers $Ra_c$ is shifted to smaller magnitudes as the Prandtl number becomes smaller.   
\end{abstract}
\maketitle

\section{Introduction}
The structure and dynamics of the viscous and thermal boundary layers near the cooled and heated plates in  thermal convection 
flow is significantly altered by the dimensionless Prandtl number $Pr\equiv \nu/\kappa$ of the working fluid \cite{Ahlers2009,Chilla2012}. For very low Prandtl 
numbers,  momentum diffusion, with the kinematic viscosity $\nu$ being the diffusion constant, is much smaller than temperature 
diffusion measured by the constant $\kappa$. Thus the thermal boundary layer is much thicker than the viscous one. This regime is of particular
interest in geo- and astrophysical convection flows. Around the solid inner core of the Earth, which is mostly an iron-nickel alloy, a liquid outer core exists
that is composed of molten metals at high pressure and $Pr\sim 10^{-1}$ \cite{King2010}. In solar and stellar convection the heat transport is supported 
by radiative processes resulting in enhanced temperature diffusion and thus  Prandtl numbers $Pr \lesssim 10^{-3}$ in the Sun \cite{Miesch2005,Hanasoge2016} 
or $Pr\lesssim 10^{-8}$ in stars \cite{Spiegel1962,Thual1992,Massaguer1991}. The majority of numerical studies of these convection flows operate at 
Prandtl (or turbulent Prandtl) numbers $Pr\sim 0.1$ to 1 in order to incorporate further physical processes of strong rotation and compressibility or
magnetic fields into the (spherical) models \cite{Miesch2000,Miesch2005}. Direct numerical simulations studies of turbulent convection below 
$Pr\lesssim 0.01$ are thus very rare. Low-Prandtl-number convection is also related 
to new methods of energy harvesting in liquid metal batteries at $Pr\sim 10^{-2}$ \cite{Sadoway2014,Kelley2014,Koellner2017}. 
In contrast, the thermal boundary layer is well embedded inside the viscous boundary layer for very large Prandtl number where temperature 
diffusion is much smaller than momentum diffusion. 

This large difference in the thicknesses of both interacting boundary layers has implications on the near-wall dynamics and eventually on a
transition to turbulence inside these boundary layers as discussed first by Kraichnan \cite{Kraichnan1962} in his landmark work. When
the Rayleigh numbers are sufficiently large at a given Prandtl number, near-wall coherent structures, such as thermal plumes for the 
temperature field and streamwise vortices for the velocity field, trigger turbulent fluctuations of both fields and change the character of the boundary
layers into a transitional one. Eventually these fluctuations fill the whole boundary layer region and the boundary layers are in a fully developed 
turbulent state. This transition to turbulence can be expected to proceed in a different way than in a standard isothermal channel flow or flat 
plate boundary layer \cite{Schlatter2014,McKeon2017} for the following two reasons: 
\begin{figure}
\centering
\vspace{0.5cm}
\includegraphics[width=0.5\textwidth]{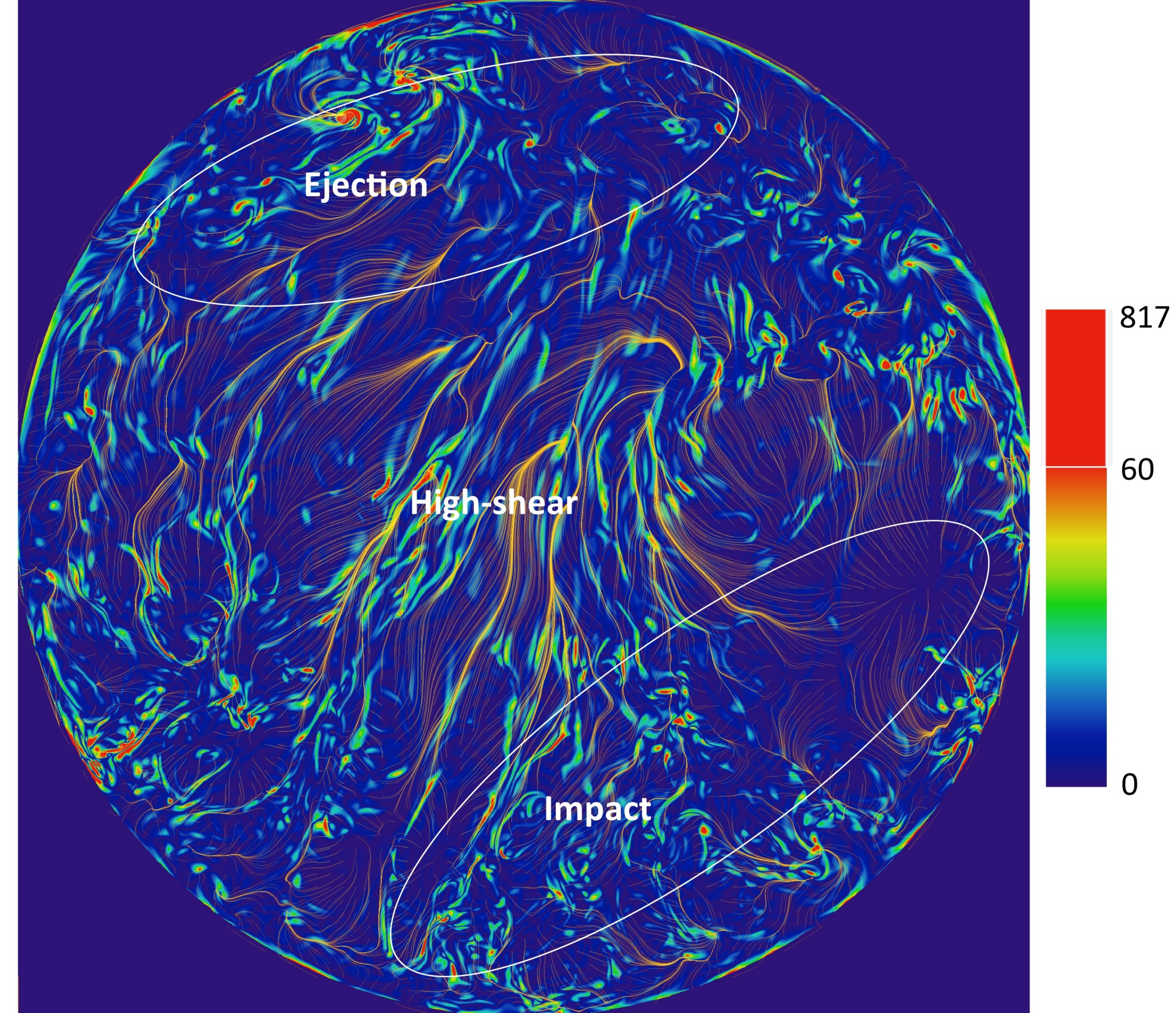}
\caption{Snapshot of the boundary layer structure for $Ra=10^7$ and $Pr=0.005$. We display the field lines of the skin friction field 
$s_i=\partial u_i/\partial x_3$ at $x_3=0$ with $i=1,2$ in yellow. They provide a two-dimensional blueprint of the velocity field near the plate. 
Colored contours correspond to the magnitude of the vorticity component  $\omega_3=\epsilon_{3jk}\partial u_k/\partial x_j$ (with $j,k=1,2$) taken 
in a horizontal plane  at $z=0.0034$ which corresponds to $2\delta_u$ (see eq. \eqref{VBL}). The magnitude of $\omega_3$ is indicated 
by the colorbar.} The three main regions of the boundary layer 
are also indicated. View is from above the boundary layer.
\label{fig1}
\end{figure}

(i)  In Rayleigh-B\'{e}nard convection (RBC) flows a well-defined canonical mean flow is missing, in particular in closed rectangular or cylindrical 
convection cells as used in all laboratory experiments. The mean flow consists then of a large-scale circulation (LSC) in form of a roll 
that breaks the symmetry and
shows a complicated temporal sloshing and oscillation dynamics \cite{Brown2006,Xia2008,Shi2012}. The LSC divides the boundary plates at the top and bottom 
into three regions, an impact region, a high-shear central region and an ejection region at approximately the diametral side of the impact region (see Fig. \ref{fig1}
for this cylindrical geometry). 
This dynamics can be partly compensated by a rotation of the coordinate frame into the instantaneous direction of this large-scale circulation flow, plane by plane 
starting from the wall \cite{Schumacher2016}. In this way a classical Reynolds-like decomposition of the turbulent fields is again possible when the analysis 
is restricted to the central region.       

(ii) The temperature field is an active scalar. Therefore thermal plumes (which are unstable fragments of the thermal boundary layer that detach
from the wall) will affect the fluid motion and enhance the velocity field fluctuations. Since these plume ridges have a typical cross section that 
corresponds to the thermal boundary layer thickness, for $Pr\gg 1$ these structures should have a small impact on the near-wall flow.
Low Prandtl number convection is characterized by fast diffusion time scales (since $\tau_d\sim \sqrt{Ra Pr}$) and much coarser plume structures.
These plume structures turn out to be very efficient drivers of fluid turbulence in the convection flow \cite{Schumacher2015} and therefore convective turbulence at very low Prandtl number is dominated by inertial fluid motion. Thus convective turbulence at low Prandtl numbers can be expected to be 
susceptible to turbulence transitions in the boundary layers at lower Rayleigh numbers than a higher-Prandtl number flow.      

With these two points in mind, in this paper we will discuss the boundary layer dynamics for $5\times 10^{-3}\le Pr\le 7\times 10^{-1}$ 
with a focus on the much less explored parameter regime of smaller Prandtl numbers. We have a comprehensive set of 
high-resolution direct numerical simulation (DNS) of three-dimensional turbulent convection flows, all obtained in exactly the same geometry, a 
closed cylindrical cell at aspect ratio $\Gamma=1$, i.e., cell diameter $D$ equals cell height $H$. The data consist of three series of Prandtl 
numbers $Pr=0.7$ for air, $Pr=0.021$ for mercury and $Pr=0.005$ for liquid sodium, the latter of which corresponds to the smallest value that 
can be obtained in controlled laboratory experiments. In the case of liquid sodium, our largest Rayleigh number value of $Ra =  10^7$ exceeds
those that have been obtained in the experiments \cite{Horanyi1999} for cells with $\Gamma\gtrsim 1$. The situation is different for convection 
in mercury at $Pr=0.021$ where Rayleigh numbers up to $Ra=10^{11}$ have been obtained in laboratory experiments which are not accessible 
in DNS. These high Rayleigh numbers correspond to Reynolds numbers of the turbulent flow of $Re\sim 10^6$ \cite{Glazier1999}

Our study summarizes global measures of transport of heat and momentum
as well as dissipation. We discuss trends of characteristic boundary layer scales with respect to Rayleigh number and compare different 
spatial intermittency measures of the velocity boundary layer. These measures require the evaluation of velocity and temperature 
derivatives inside the boundary layers which are not accessible in laboratory experiments, in particular for larger Rayleigh numbers.
 Their extrapolation to higher Rayleigh numbers allows us to predict a range of Rayleigh numbers beyond which the transition to turbulent boundary layers 
might be present and convection crosses over into Kraichnan's ultimate regime \cite{Kraichnan1962}. This transition is expected to be subcritical just 
as for isothermal boundary layers and thus a range of Rayleigh numbers for the transition can be expected rather than a sharp threshold 
\cite{Schlatter2014}. Our ranges of Rayleigh numbers  follow, however, from the uncertainty of the extrapolation.
We discuss the numerical method in the next section. This is followed by our analysis of global transport and of boundary layer scalings. We then summarize our results and discuss possible future directions. 

\begin{figure}
\centering
\vspace{0.5cm}
\includegraphics[width=0.47\textwidth]{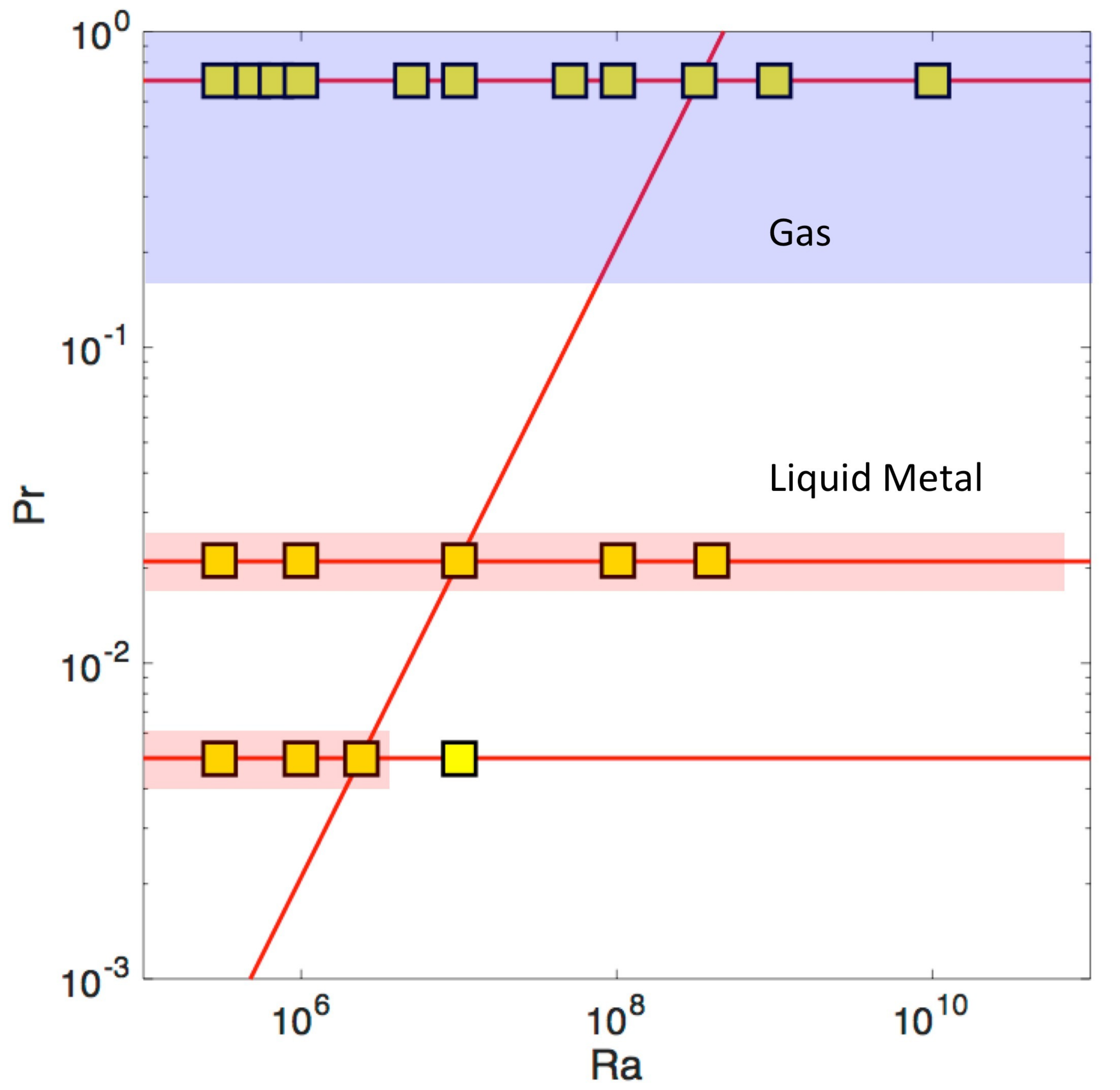}
\caption{Simulation data record plotted in the parameter plane which is spanned by $Ra$ and $Pr$. The blue shaded area
of the parameter plane for $Pr>0.16$ is accessible for gases and/or binary gas mixtures \cite{Liu1997}.  The red shaded areas 
indicate the range of Rayleigh numbers which are covered by laboratory experiments in cases of liquid mercury and liquid sodium for cells with 
$\Gamma\sim 1$. These are the experiments by Glazier et al. \cite{Glazier1999} up to $Ra=8\times 10^{10}$ for mercury and 
Horanyi et al. \cite{Horanyi1999} up to $Ra=5\times 10^6$ for sodium. Three of our simulations are on a line of constant Grashof
number $Gr=Ra/Pr$ which has been discussed in \cite{Schumacher2015}. The constant $Gr$ data points are connected by the inclined red line.
A summary of further low-Prandtl-number experiments is found in \cite{Scheel2016}.}
\label{fig2}
\end{figure}

\section{Simulations}
We solve the three-dimensional equations of motion in the Boussinesq approximation. They couple the velocity field $u_i(x_j,t)$ with 
the temperature $T(x_j,t)$. The indices $i, j, k=1,2,3$, and the Einstein summation convention will be used in this work. The equations are made dimensionless by using
height of the cell $H$, the free-fall velocity $U_f=\sqrt{g \alpha \Delta T H}$ and the imposed temperature difference $\Delta T$. The equations
contain the three control parameters: the Rayleigh number $Ra$, the Prandtl number $Pr$ and the aspect ratio $\Gamma=D/H=1$ with the 
cell diameter $D$. The equations are given by
\begin{align}
\label{ceq}
\frac{\partial u_i}{\partial x_i}&=0\,,\\
\label{nseq}
\frac{\partial  u_i}{\partial  t}+u_j \frac{\partial u_i}{\partial x_j}
&=-\frac{\partial p}{\partial x_i}+\sqrt{\frac{Pr}{Ra}} \frac{\partial^2 u_i}{\partial x_j^2}+  T \delta_{i3}\,,\\
\frac{\partial  T}{\partial  t}+u_j \frac{\partial T}{\partial x_j}
&=\frac{1}{\sqrt{Ra Pr}} \frac{\partial^2 T}{\partial x_j^2}\,,
\label{pseq}
\end{align}
where
\begin{equation}
Ra=\frac{g\alpha\Delta T H^3}{\nu\kappa}\,,\;\;\;\;\;\;\;\;Pr=\frac{\nu}{\kappa}\,.
\end{equation}
The variable $g$ stands for the  acceleration due to gravity, $\alpha$ is the thermal expansion coefficient, $\nu$ is the kinematic viscosity, and 
$\kappa$ is thermal diffusivity. No-slip boundary conditions for the fluid are applied at all walls. The side walls are thermally insulated and the 
top and bottom plates are held at constant dimensionless temperatures $T=0$ and 1, respectively. 

\begin{table*}
\begin{center}
\begin{tabular}{rcccccccccc}
\hline\hline
Run  &  $Pr$  &  $Ra$  &   $N_e$  &   $\;\;\;N\;\;\;$  & $\;\;\;N_{bl}\;\;\;$ &  $Nu$     &      $Re$  &  $Re_{\tau}$ &  Snapshots &  Runtime \\
\hline
1  &  0.7 &  $3\times 10^5$ &   3,520  & 11  &  27 &  $5.80\pm 0.03$  &  $116\pm 1$     & --  & 401   &    695\\                      
2  &  0.7 &  $5\times 10^5$ &   3,520  & 11  &  24 &  $6.90\pm 0.13$  & $151\pm 2$	    & --  & 401	&	682\\
3  &  0.7 &  $7\times 10^5$ &   3,520  & 11  &  20 &  $7.78\pm 0.05$  & $179\pm 1$	    & --  & 407	&	682\\
4  &  0.7 &  $10^6$  &  30,720            &   7  &   25 &  $8.65\pm 0.06$  & $214\pm 6$	    & 5  & 300	&	542\\
5  &  0.7 &  $5\times 10^6$  & 30,720 &  7   &  17 &  $13.79\pm 0.17$ & $483\pm 1$	    & 7  & 340    &	459\\
6  &  0.7 &  $10^7$  & 30,720             & 11   &  21 &  $16.77\pm 0.01$ & $675\pm 3$	    & 8  & 230  &	177\\
7  &  0.7 &  $5\times 10^7$  & 30,720& 13   &  18 &  $25.8\pm 0.3$   & $1,490\pm 40$  & 12 &192	  &	 93\\
8  &  0.7 &  $10^8$  & 256,000           & 11   &  27 &  $31.4\pm 1.3$   & $2,070\pm 60$  & 14 &84    &	 79\\		 
9  &  0.7 &  $10^9$  & 875,520           & 11   & 18  &  $63\pm 4$         & $6,240\pm 140$ & 24&92	  &	 72\\
10&  0.7  & $10^{10}$ &2,374,400      &  9   &   8  &  $127\pm 6$       & $19,300\pm 900$	& 48& 41	  &	 48\\
\\
11&  0.021 & $3\times 10^5$ &  256,000   & 7  & 54  &  $4.29\pm 0.12$ &   $1,830\pm 30$   & 18 &	68	&	92 \\
12&  0.021 & $10^6$  & 256,000               & 9   & 48  &  $5.43\pm 0.03$ &   $3,030\pm 40	$    &  24 &150  &	131 \\
13&  0.021 & $10^7$  & 875,520               & 11  & 26  &  $10.11\pm 0.05$ & $8,450\pm 100$   &  35 & 206  &	 29 \\		 
$^{\ast}$14&  0.021 & $10^8$  & 2,374,400            & 13  & 18  &  $19.1\pm 1.3$  &  $22,900\pm 300$  & 48  &   27  &	 12 \\
$^{\ast}$15&  0.021 & $4\times 10^8$ & 6,272,000 & 11  & 42  &  $30.8\pm 1.7$  &  $46,000\pm 600$  & 76  &  75  &	 6  \\
\\
16&  0.005 & $3\times 10^5$ & 256,000         &  9 &  43  &  $3.26\pm 0.02$ & $4,620\pm 20$	& 43 & 29	& 30 \\
17&  0.005 & $10^6$  & 875520                      &  9 &  39  &  $4.45\pm 0.07$ & $8,290\pm 40$	 & 49 &  63	& 39 \\
18&  0.005 & $2.38\times 10^6$ & 2,374,400 & 11 &  30  &  $5.66\pm 0.09$ & $12,800\pm 120$  &	65 & 25 	& 8   \\
$^{\ast}$19&  0.005 & $10^7$  & 5,644,800                  &  9  &	 148 &  $8.0\pm 0.7$   &  $20,800\pm 160$  &	54 & 19 &  11 \\
\hline\hline
\end{tabular}  
\caption{Parameters of the different spectral element simulations. We display the Prandtl number $Pr$, the 
Rayleigh number $Ra$, the number of spectral elements $N_e$, the polynomial order of the expansion in each 
of the three space directions, the number of mesh cells inside the thermal boundary layer $N_{bl}$, the Nusselt 
number $Nu$,  the Reynolds number $Re$, the friction Reynolds number $Re_{\tau}$ the number of statistically independent snapshots, and the total runtime 
in units of the free-fall time $T_f=H/U_f$. The runs with one asterisk are either new or have been run for a longer time to gather
a better statistics in comparison to \cite{Scheel2016,Schumacher2016}.}
\label{Tab1}
\end{center}
\end{table*}
The equations are numerically solved by the Nek5000 spectral element method package which has been adapted to our problem 
\citep{nek5000,Scheel2013}. The cylindrical cell is composed of spectral elements. On each element, the turbulent fields are expanded in 
Lagrangian interpolation polynomials of order $N$ in each space direction. This sums up to more than 6 million spectral elements and
approximately 10 billion mesh cells (see Tab. \ref{Tab1}) which require more than half a million Message Passing 
Interface tasks at the Blue Gene/Q supercomputers JUQUEEN in J\"ulich (Germany) and Mira at Argonne National Laboratory (USA). 
As seen in the table, a particular challenge are the largest Rayleigh number runs at the lowest Prandtl number which require small integration time steps 
due to the high diffusivity of the temperature field. For example, run 19 consumed 22 million processor core hours for runs with 131,072 
Message Passing Interface tasks while advancing for 11 free-fall times $T_f$. It is clear that a longer time series would be desirable.

Fig. \ref{fig2} summarizes all simulation data sets and relates them to ranges of Rayleigh numbers that have been reached in the laboratory 
in a comparable geometry and Prandtl number. The figure displays also the range of Prandtl numbers that are accessible to gases and binary 
gas mixtures and thus to optical imaging technology.   
\begin{figure}
\centering
\vspace{0.5cm}
\includegraphics[width=0.32\textwidth]{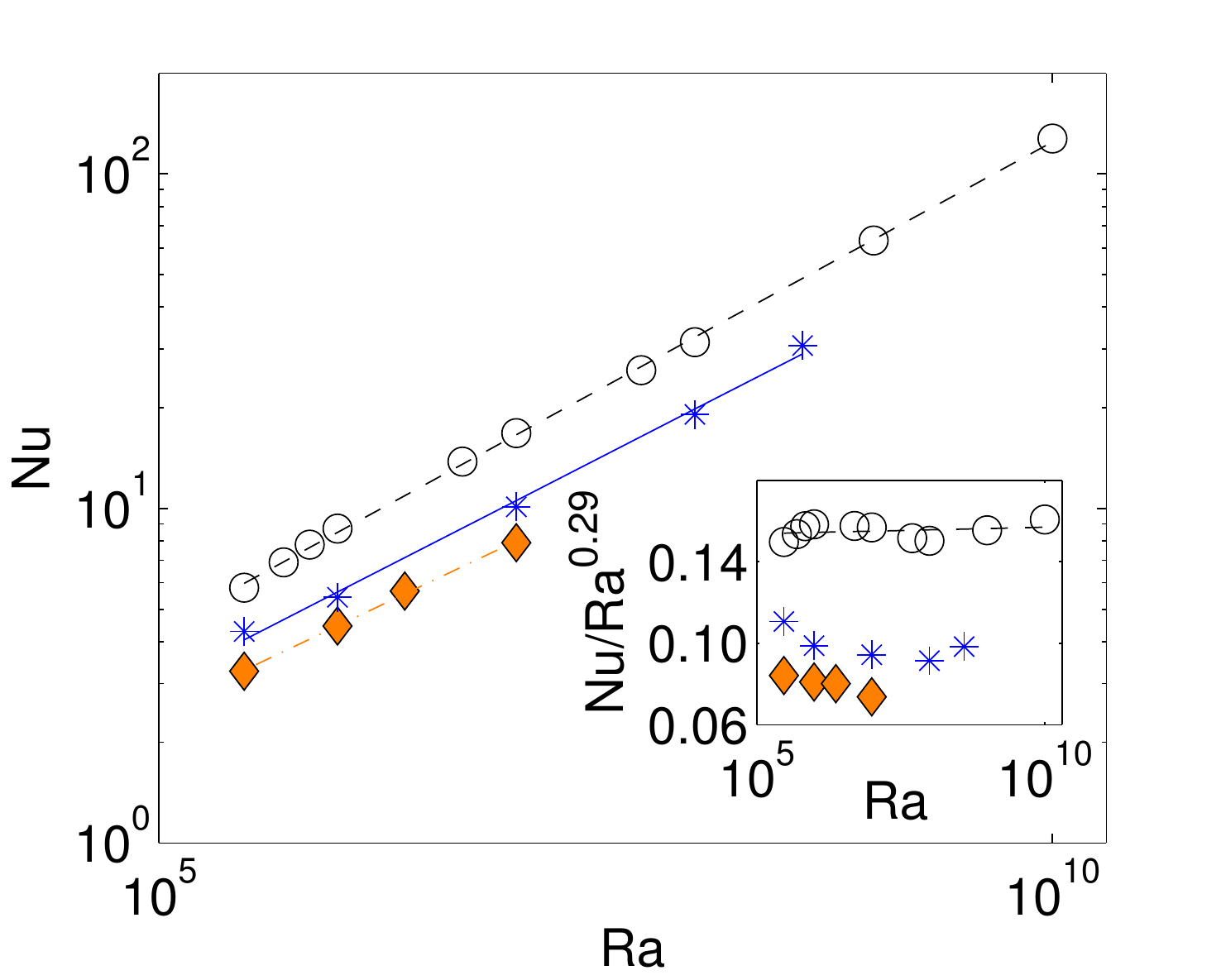}
\includegraphics[width=0.32\textwidth]{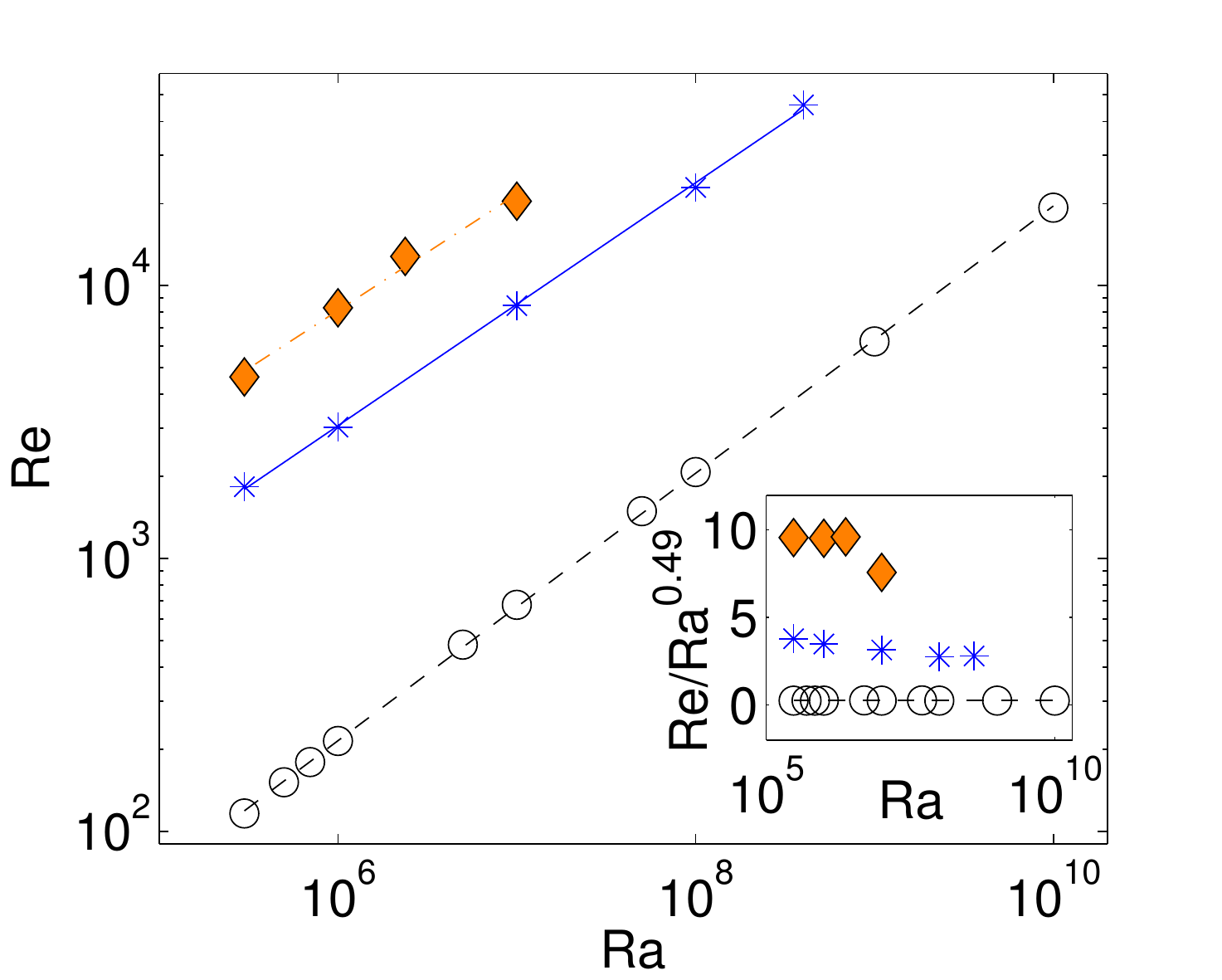}
\includegraphics[width=0.33\textwidth]{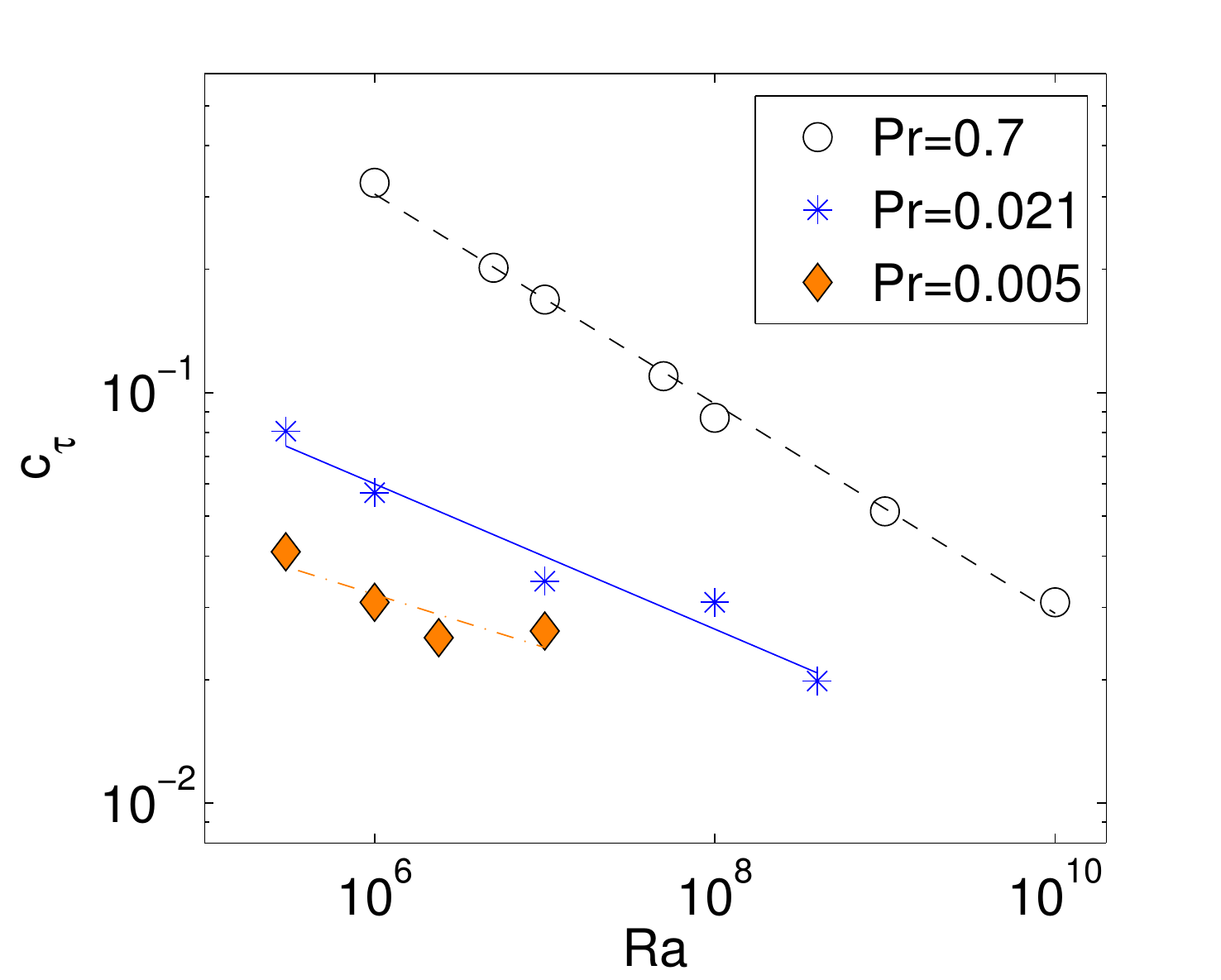}
\caption{Global turbulent transport and dissipation laws versus Rayleigh number. Left: Scaling laws of turbulent heat transfer, $Nu(Ra)$,  
for the three series of DNS. The inset compensates all plots with respect to the scaling exponent for convection in air. Middle: 
Scaling laws of turbulent momentum transfer, $Re(Ra)$. The inset replots data  and compensates the series with respect to the scaling 
for convection in air. Right: Skin friction coefficient $c_{\tau}$ determined over the inner plate section with $r\le 0.3$ denoted with an index $b$ as 
a function of the Rayleigh number. Open circles correspond  to  $Pr=0.7$, asterisks to  $Pr=0.021$ and filled diamonds to $Pr=0.005$
in all panels.}
\label{fig3}
\end{figure}

\section{Global turbulent transport and dissipation}
In response to the three input parameters $Ra$, $Pr$ and $\Gamma$,  turbulent heat and momentum fluxes are established in an RBC flow. 
The turbulent heat transport is determined by the dimensionless Nusselt number which is given by 
\begin{equation}
Nu=1+\sqrt{Ra Pr}\langle  u_3  T\rangle_{V,t}\,.
\label{Nusselt}
\end{equation}
Expression $\langle\cdot\rangle_{V,t}$ stands for a combined volume and time average. The mean thickness of the 
thermal boundary layer is given by 
\begin{equation}
\delta_T=\frac{1}{2 Nu}\,.
\label{TBL}
\end{equation}
The left panel of Fig. \ref{fig3} summarizes the global turbulent transport laws of heat obtained for the three series 
of DNS. We fit the data to the following scaling laws:
\begin{equation}
\label{NuRa}
Nu = \left \{ \begin{array}{lcl}
                               (0.13 \pm 0.02) Ra^{0.26 \pm 0.01}  & : & Pr=0.005\\
                               (0.13 \pm 0.04) Ra^{0.27 \pm 0.01}  & : & Pr=0.021\\
                               (0.15 \pm 0.01) Ra^{0.29 \pm 0.01}  & : & Pr=0.700
                 \end{array} \right.
\end{equation}
A decrease of the scaling exponent from 0.29 for convection in air to 0.26 in sodium is observed. Theoretical predictions for the 
scaling exponents at different Prandtl numbers are found in the range between 1/3 and 1/4. The exponent of 1/3 would follow from a 
model that describes thermal boundary layers as being marginally stable, an idea which goes back to Malkus \cite{Malkus1954}.  The scaling 
theory of Shraiman and Siggia \cite{Shraiman1990} builds on a turbulent boundary layer and gives an exponent of  2/7 for $Pr>1$. Our 
exponent of 0.29 for $Pr=0.7$ is consistent with 
the Grossmann-Lohse theory that however predicts no pure algebraic scaling \cite{Ahlers2009,Stevens2013}. An exponent of 1/4 follows from a joint 
balancing of inertial and buoyancy term in the momentum balance together with a balance of advection and diffusion terms in the 
advection-diffusion equation for temperature \cite{Jones1976}. It is also the result of an asymptotic model for steady inertial convection in 
two dimensions by Busse and Clever which is formulated for the limit of very small Prandtl numbers \cite{Busse1981}. In both models 
\cite{Jones1976,Busse1981} of low-Prandtl convection, it is assumed that the fluid turbulence is highly inertial. The compensated 
plots in the insets of both panels show that this scaling is not perfect, as it has been found in many other studies of thermal convection.
When the data of run 19 are excluded from the fit, the exponent for the sodium series becomes $0.265\pm 0.01$ as given in \cite{Scheel2016}. 
The exponents are found in the same range as laboratory experiments in air \cite{Chilla2012} and in mercury or gallium 
\cite{Cioni1997,Glazier1999,Aurnou2015} as well as in sodium \cite{Horanyi1999}.  
\begin{figure}
\centering
\includegraphics[width=0.9\textwidth]{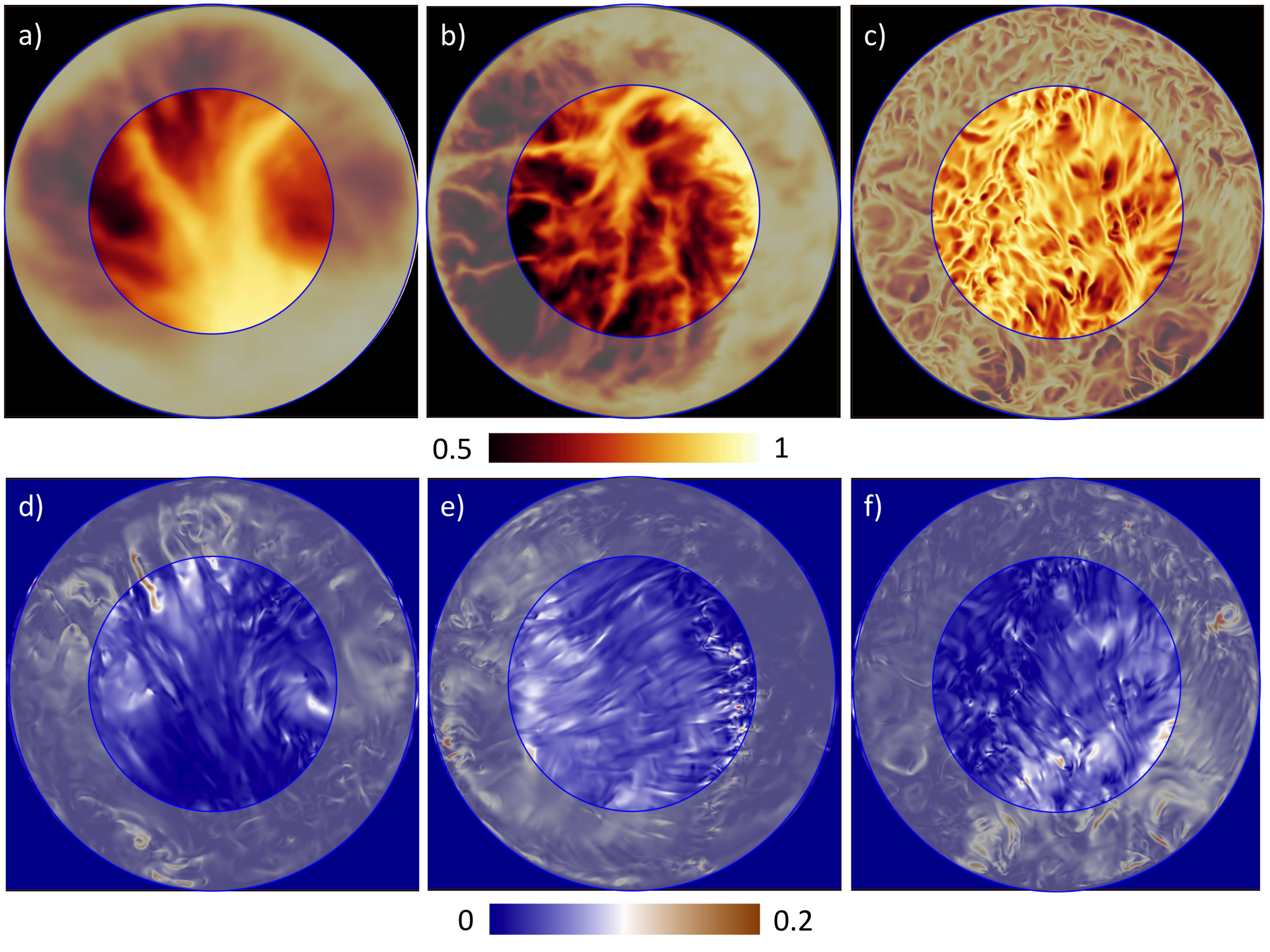}
\caption{Snapshots of the temperature field in the top row and of the wall shear stress field in the bottom row. 
(a,d) Run 19 at $Ra=10^7$ and $Pr=0.005$. (b,e) Run 14 at $Ra=10^8$ and $Pr=0.021$. (c,f) Run 10 at $Ra=10^{10}$ 
and $Pr=0.7$. All three runs have a comparable Reynolds number $Re\approx 2\times 10^4$ as seen in Tab. \ref{Tab1}.  The interior 
plate section is always highlighted. The corresponding color scales are the same in the three cases. The cross section planes 
for $T$ are taken at $z=\delta_T/2$ with $\delta_T=1/(2 Nu)$ being the thermal boundary layer thickness. The ratio of the thicknesses 
is approximately $1:7:16$ when convection in sodium, mercury, and air are compared. The instantaneous wall shear stress field
can be derived from Eqns.~(\ref{cfsg}) and  (\ref{flux4}) as $\tau_w(x_1,x_2)=Gr^{-1/2} |\partial_3 u_i|$ with Grashof number $Gr=Ra/Pr$ and 
summation over $i=1,2$. View is from below in all figures.}
\label{fig4}
\end{figure}

The turbulent momentum transport is expressed by the Reynolds number $Re$ which is defined as
\begin{equation}
Re=u_{rms}\sqrt{\frac{Ra}{Pr}}\;\;\;\;\text{with}\;\;\;\;u_{rms} =\sqrt{\langle u_i^2\rangle_{V,t}}\,,
\label{Reynolds}
\end{equation}
The mean thickness of the viscous boundary layer is frequently given by Blasius-type relation 
\begin{equation}
\delta_u=\frac{a}{\sqrt{Re}}\,,
\label{VBL}
\end{equation}
where $a<1$ is a free parameter (here $a=1/4$ as in \cite{Grossmann2001}). 
The following scaling laws can be derived from our data (see mid panel of Fig. \ref{fig3}):      
\begin{equation}
\label{ReRa}
Re = \left \{ \begin{array}{lcl}
                               (21 \pm 9) Ra^{0.43 \pm 0.03}  & : & Pr=0.005\\
                               (6.5 \pm 0.6) Ra^{0.45 \pm 0.01}  & : & Pr=0.021\\
                               (0.24 \pm 0.01) Ra^{0.49 \pm 0.01}  & : & Pr=0.700
                 \end{array} \right.
\end{equation}
Again, the trend is very systematic with an exponent that decreases for decreasing Prandtl number and with a prefactor that is 
systematically enhanced confirming the high level of developed fluid turbulence in the cell. If we exclude run 19 from the fit 
to the data for the series of convection in sodium, an exponent of $0.49\pm 0.01$ would follow \cite{Scheel2016}.
For the turbulent momentum transfer 
fewer (direct) measurements are available. In nearly all cases the temperature signal is used to determine a Reynolds number
on the basis of a characteristic frequency, e.g. in Ref. \cite{Takeshita1996}. Our scaling exponent for mercury agrees with the experiments 
\cite{Takeshita1996} as does that for air \cite{Chavanne2001}.

Finally it is worth noting that we observe fairly constant slopes for all Reynolds versus Rayleigh data the mid panel of Fig. \ref{fig3}, 
and particularly for Pr=0.7 which covers Reynolds numbers that span two orders of magnitude. For all $Re$ the bulk flow is turbulent, and 
the boundary layer that can be characterized by an increasing level of turbulent fluctuations as $Ra$ increases, but is still always below 
the transition to a fully turbulent boundary layer.

Most of the kinetic energy is dissipated in the boundary layers. A global measure of the dissipative losses in shear flows is the 
skin-friction coefficient (see e.g. \cite{Schlichting2000}) which is given in the present flow \cite{Scheel2016} in dimensionless form by
\begin{equation}
c_{\tau}=\frac{2 \langle\tau_w\rangle_{b,t}}{u_{rms}^2}\,.
\label{cfsg}
\end{equation}
Here, $\langle\tau_w\rangle_{b,t}$ is the mean wall shear stress. This stress field is connected with the friction velocity 
$u_{\tau}$ via $\langle\tau_w\rangle_{b,t} = u_{\tau}^2$. Following \cite{Schumacher2016}, the friction velocity is given by     
\begin{equation}
\label{flux4}
u_{\tau}(x_3=0)= \sqrt[4]{\frac{Pr}{Ra}}\, \Biggl\langle\left(\Bigg\langle\frac{\partial u_i}{\partial x_3} 
\Bigg\rangle^2_{b}\right)^{\frac{1}{4}} \Biggl\rangle_t\,,
\end{equation}
with $i=1,2$. 
Fig. \ref{fig3} (right) summarizes the results for $c_{\tau}$ for all simulation data. The following algebraic power laws have been fitted to the 
recent data sets. 
\begin{equation}
\label{CtauRa}
c_{\tau} = \left \{ \begin{array}{lcl}
                               (0.2 \pm 0.1) Ra^{-0.12 \pm 0.05}  & : & Pr=0.005\\
                               (0.7 \pm 0.2) Ra^{-0.18 \pm 0.02}  & : & Pr=0.021\\
                               (10 \pm 1) Ra^{-0.26 \pm 0.01}  & : & Pr=0.700
                 \end{array} \right.
\end{equation}
Note that this calculation method differs from the one in \cite{Scheel2016} which alters the scaling laws slightly. Without run 19 the power law fit for 
the series at $Pr=0.005$ is  $c_\tau=(0.78 \pm 0.01) Ra^{-0.23 \pm 0.01}$. We have chosen
the present method in order to conduct the analysis consistently with the subsequent sections where the logarithmic law of the wall will be discussed.  
While a laminar isothermal flat plate boundary layer would approximately scale as $c_{\tau}\sim Ra^{-1/4}$, a turbulent boundary layer would follow
$c_{\tau}\sim Ra^{-1/10}$ for larger Reynolds numbers if we simply translate $Re \sim \sqrt{Ra}$ \cite{Schlichting2000}.   
We observe that the magnitude of the scaling exponent for air coincides with the one for a laminar boundary layer and that this exponent increases for sodium to almost a fully turbulent boundary layer scaling. This crossover is caused by the increasing discrepancy between the thicknesses of thermal and 
viscous boundary layers as $Pr$ falls and the fact that the  temperature is an active scalar that affects the fluid motion near the walls. In Fig. \ref{fig4}, we plot three horizontal cuts 
well inside the corresponding thermal boundary layer for runs 10, 14 and 19. All three simulations result in a comparable $Re\approx 20,000$. Bright 
contours stand for the detachment zones of thermal plumes. Coarse temperature field contours (see left panel for sodium) are then in line with regions across which wall shear stress filaments vary only moderately. This can be seen in the bottom row of the same figure. The region containing 
 detaching plumes is largest in the sodium case and results in a reduced drop in the friction coefficient with respect to $Ra$. 

\begin{figure}
\centering
\includegraphics[width=0.32\textwidth]{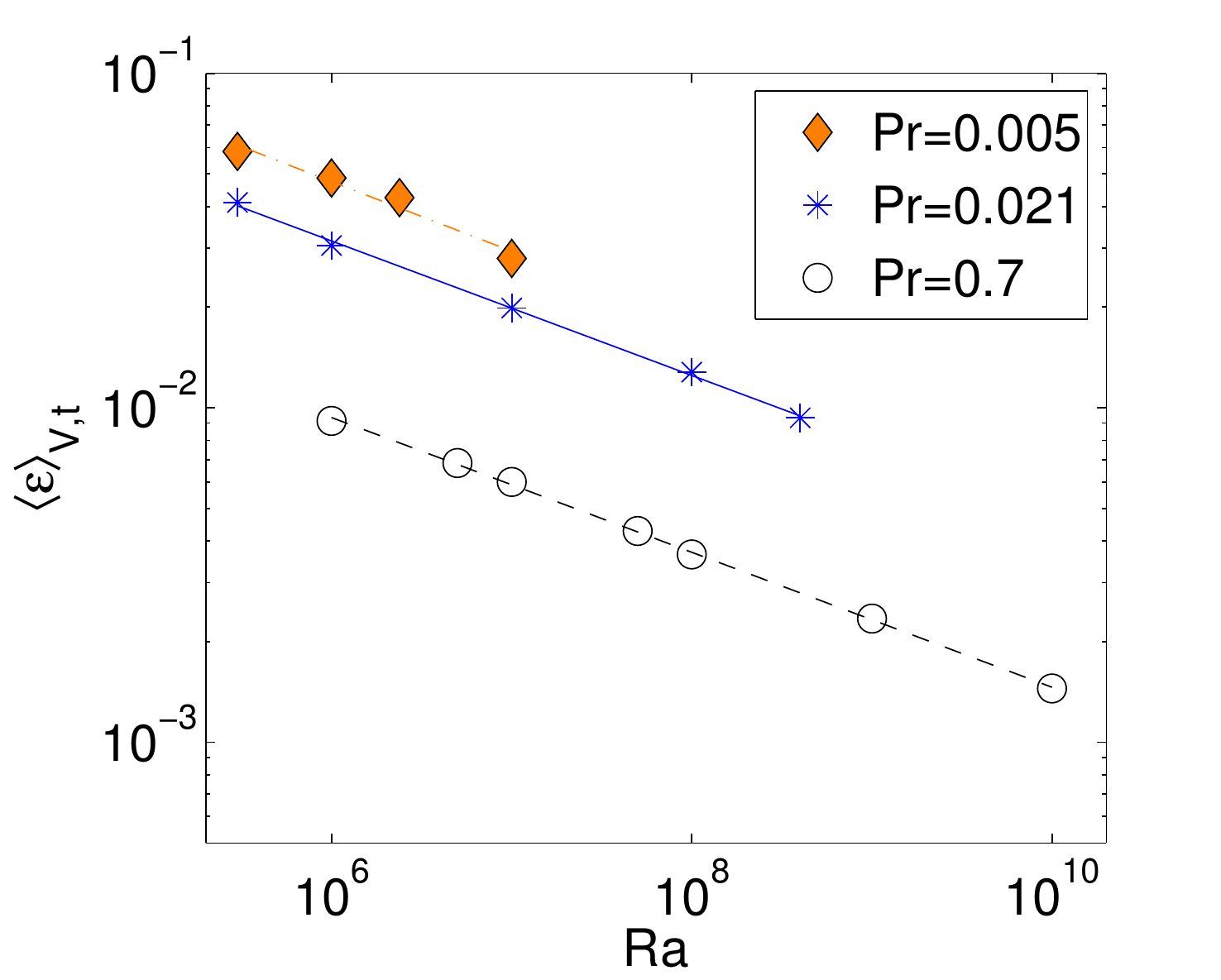}
\includegraphics[width=0.32\textwidth]{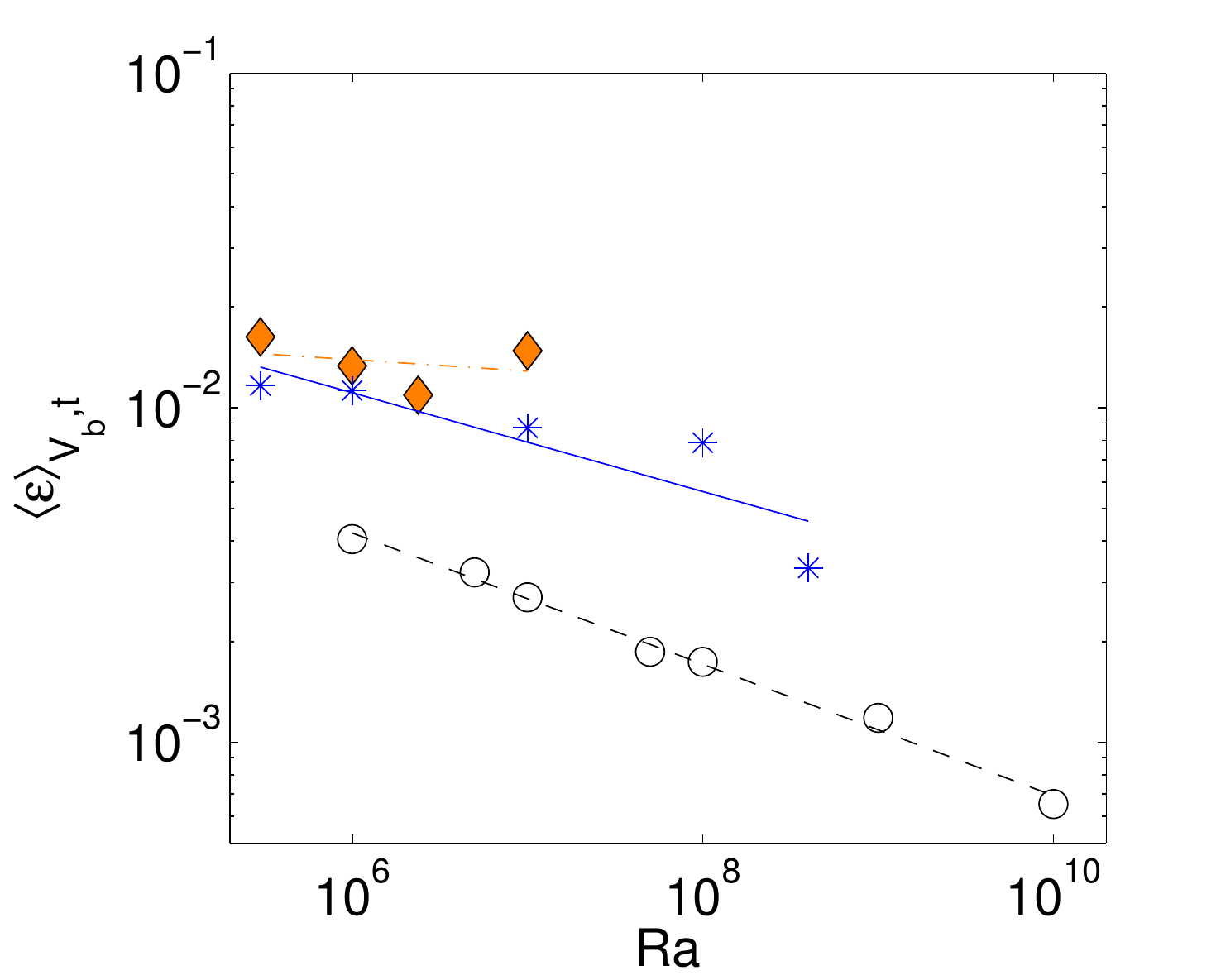}
\includegraphics[width=0.32\textwidth]{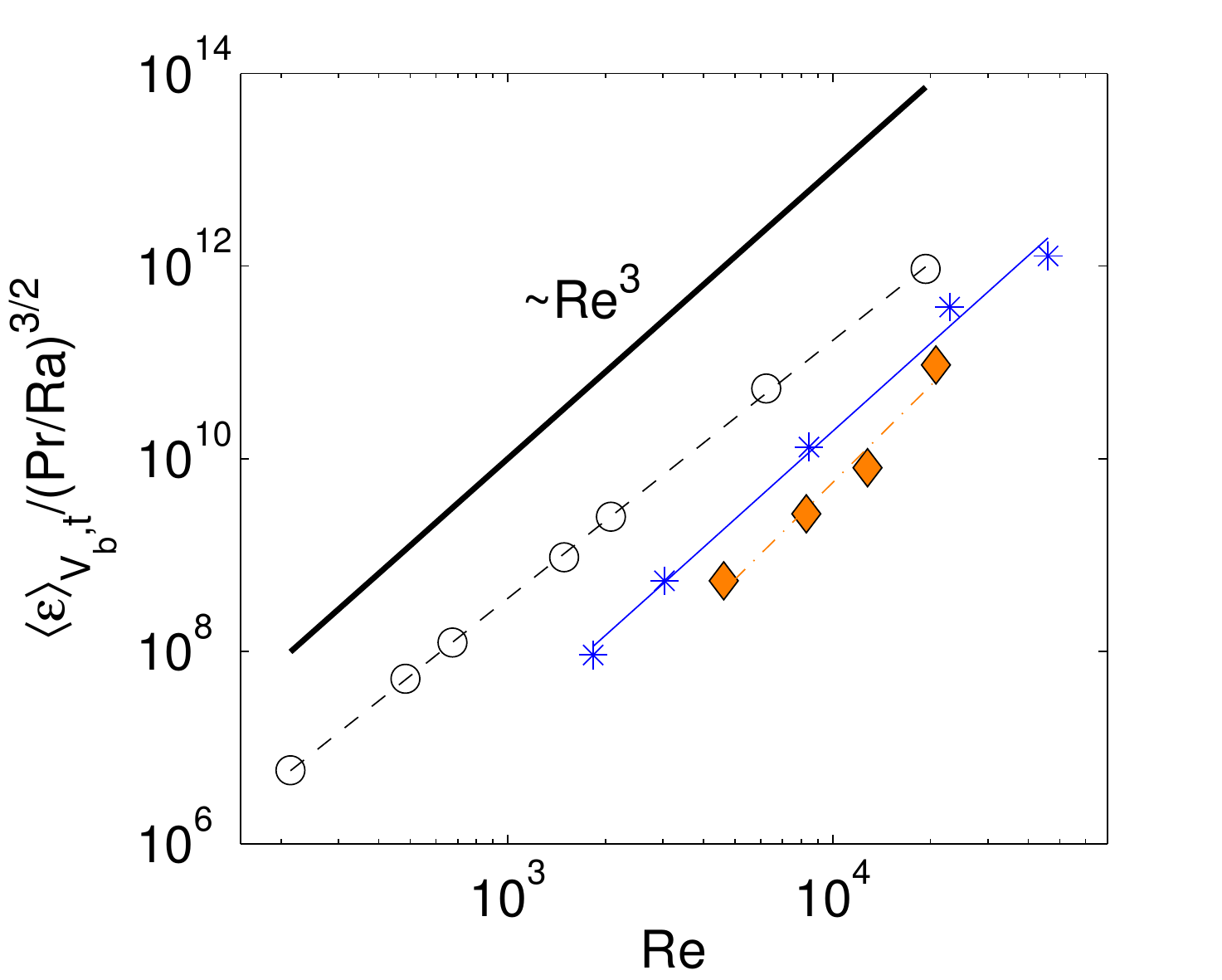}
\caption{Mean kinetic energy dissipation rate as a function of the Rayleigh and Prandtl numbers. Left: Mean kinetic energy dissipation rate 
$\langle\epsilon\rangle_{V,t}$ obtained in the full cylinder volume $V$. The scaling laws that were fitted to the data results in the following 
scaling exponents: $-0.20\pm 0.01$ for $Pr=0.7$ (open circles), $-0.20\pm 0.01$ for $Pr=0.021$ (asterisks) and $-0.21\pm 0.03$ for $Pr=0.005$ (diamonds). 
Middle: Mean dissipation rates in a subvolume $V_b$ centered in the bulk away from all walls, $\langle\epsilon\rangle_{V_b,t}$. 
This results in the following scaling exponents: $-0.19\pm 0.01$ for $Pr=0.7$ (open circles), $-0.15\pm 0.05$ for $Pr=0.021$ (asterisks) and $-0.03\pm 0.08$ for $Pr=0.005$ (diamonds). Without run 19 the exponent is  $-0.19\pm 0.02$ for $Pr=0.005$. The subvolume in the bulk takes 22\% of the total cell volume. Right: 
Replot of the mean kinetic energy dissipation rates in the bulk in versus Reynolds number to demonstrate ansatz (\ref{ediss2}).}
\label{fig5}
\end{figure}

Another global dissipation measure is the volume mean of the kinetic energy dissipation rate. The corresponding field is defined as
\begin{equation}
\label{ediss}
\epsilon(x_k,t) = \frac{1}{2}\sqrt{\frac{Pr}{Ra}} \left(\frac{\partial u_i}{\partial x_j}+ \frac{\partial u_j}{\partial x_i}\right)^2\,. 
\end{equation}
The kinetic energy balance in a statistically stationary regime gives rise to an exact relation for the mean kinetic energy dissipation rate
\cite{Scheel2013}
\begin{equation}
\label{ediss1}
\langle\epsilon\rangle_{V,t} = \frac{Nu-1}{\sqrt{Ra Pr}}\,. 
\end{equation}
In Fig. \ref{fig5}(a), we plot the mean kinetic energy dissipation rate for the three DNS data series obtained from a full volume and time average of Eqn. (\ref{ediss}). 
\begin{figure}
\centering
\includegraphics[width=0.45\textwidth]{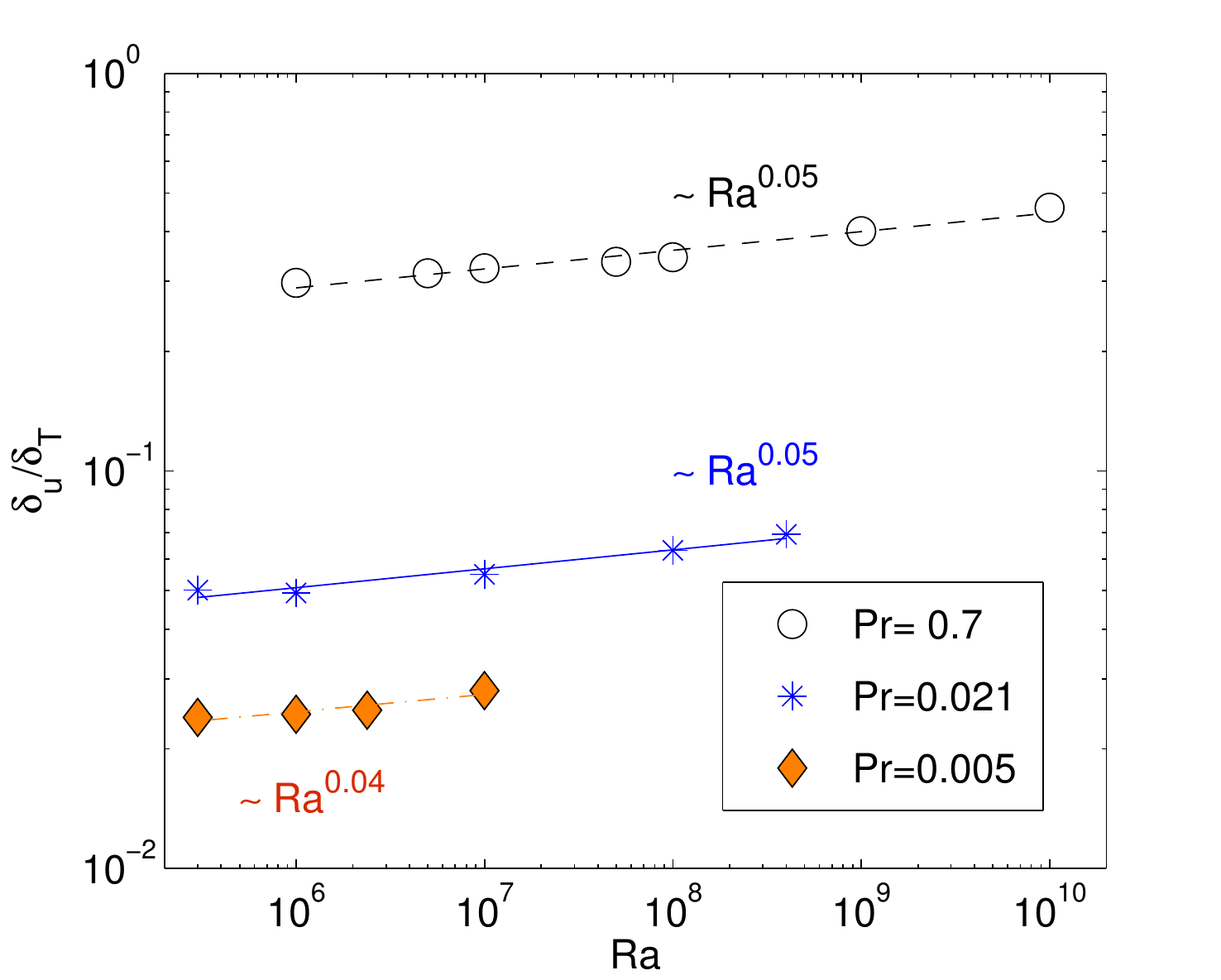}
\caption{Ratio of mean boundary layer thicknesses $\delta_u/\delta_T$ versus Rayleigh number $Ra$. Symbols are given in the legend.}
\label{fig6}
\end{figure}
It is observed that the power law exponent of the scaling of the dissipation rate with respect to Rayleigh number is about the same for all
series and varies between 0.20 and 0.21. At a given $Ra$, the magnitude of mean kinetic energy energy dissipation is enhanced as 
$Pr$ decreases which is in line with a growing relevance of  the inertial effects of the fluid turbulence. Following Grossmann and Lohse 
\cite{Grossmann2000}, we decompose the mean kinetic energy dissipation rate into a boundary layer (bl) and bulk (bu) contribution. 
Dimensional analysis gives the following Reynolds number dependencies of both contributions of the mean dissipation rate with a physical 
dimension that is indicated here with a tilde 
\begin{equation}
\label{ediss2}
\langle\tilde\epsilon\rangle_{V,t} = \langle\tilde\epsilon_{bl}\rangle_{V,t}+ \langle\tilde\epsilon_{bu}\rangle_{V,t}
=c_1\frac{\nu^3}{H^4}Re^{\frac{5}{2}}+c_2\frac{\nu^3}{H^4}Re^{3}\,. 
\end{equation}
In Fig.~\ref{fig5}(b) the mean kinetic energy dissipation rate in the bulk of the cell is displayed. The combined volume-time 
average is taken in a cylindrical subvolume $V_b$ in the center of cell that occupies about 22\% of the total volume $V$ and is sufficiently 
far away from all boundary layers for all $Ra$ and $Pr$, including those at the sidewalls. It is seen that the dissipation rate magnitude is 
significantly smaller in the bulk, by a factor of two or larger. The magnitude of the power law exponents that were fitted to the data are 
slightly smaller and decrease with decreasing $Pr$. If we take the Reynolds number based on the free-fall velocity, 
$Re_f= U_f H/\nu=\sqrt{Ra/Pr}$, eq.~(\ref{ediss2}) translates to 
\begin{equation}
\label{ediss3}
\frac{\langle\epsilon\rangle_{V,t}}{\sqrt{(Pr/Ra)^3}} = c_1Re^{\frac{5}{2}}+c_2 Re^{3}\,.
\end{equation}
Figure~\ref{fig5}(c) replots the mean dissipation rates in the bulk in correspondence with eq.~(\ref{ediss3}) and reproduces a cubic scaling 
with the Reynolds number that is taken from Tab.~\ref{Tab1}.

Following from Tab. II of Ahlers et al. \cite{Ahlers2009}, our data correpond to regime II$_{\ell}$
for which kinetic energy dissipation in the bulk and thermal dissipation in the boundary layer dominate, respectively. In this regime, pure scaling laws 
$Nu\sim Ra^{1/5} Pr^{1/5}$ and $Re\sim Ra^{-2/5} Pr^{-3/5}$ would follow. Our series for the lower $Pr$ trends towards this scaling.

Is is also interesting to plot  the ratio of the viscous and thermal boundary layer thicknesses here, which is displayed in Fig. \ref{fig6}.  
The ratio is smaller than one for all runs displayed. There is, however, a fairly significant decrease in the ratio as Prandtl number decreases. 
The resulting scaling with respect to Rayleigh number is a consequence of the obtained scaling laws $Nu(Ra,Pr)$ and $Re(Ra,Pr)$ in 
combination with the definitions (\ref{TBL}) and (\ref{VBL}). The scaling exponent of the thickness ratio remains thus nearly unchanged for decreasing 
Prandtl number. This is a consequence of the fact that the Nusselt and Reynolds scaling exponents both decrease in magnitude as Prandtl number 
decreases (see also eqns. (\ref{NuRa}) and (\ref{ReRa})). For other definitions of thermal and viscous boundary layer thicknesses, and a more 
systematic comparison for Rayleigh and Prandtl numbers, see also refs.~\cite{Scheel2016, Scheel2014}.

\section{Transition to turbulence in the viscous boundary layer} 
\subsection{Method I: Friction Reynolds number from logarithmic law of wall}
In ref. \cite{Schumacher2016}, we showed that the viscous boundary layer is characterized by a favorable pressure gradient 
which is connected with the LSC flow in the cell. Since this LSC is twisted and obeys a complicated three-dimensional dynamics, in particular
an ongoing change of the orientation, a rotation of the frame of reference into the mean orientation of the LSC at each plane of constant 
$x_3$ was suggested. This plane-by-plane rotation, which is done for each snapshot separately, removes the interior twist of the roll, by compensating for the orientation changes and allows 
us to define  streamwise and spanwise directions similar to plane wall-bounded shear flows. Details of the transformation can be found in the 
appendix. Together with the transformation $\hat{x}_i=R_{ij} x_j$, a corresponding transformation for the velocity components is given as
$\hat{u}_i=R_{ij} u_j$.

\begin{figure}
\centering
\includegraphics[width=0.45\textwidth]{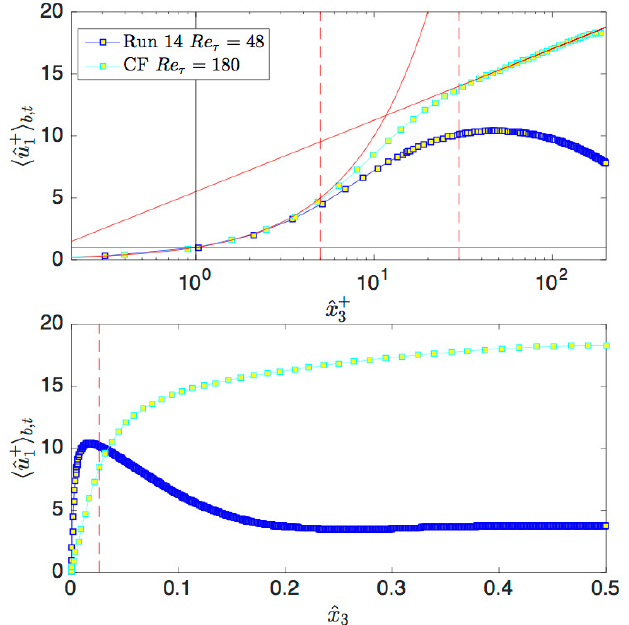}
\caption{Mean streamwise velocity profile for a turbulent RBC flow (run 14). Top: Plot of the data in inner wall units for comparison with the 
logarithmic law of the wall for a turbulent channel flow \cite{Alamo2003} at $Re_{\tau}=180$. Profiles have to intersect $(\hat{x}_3^+, \hat{u}_1^+)=(1,1)$
which is indicated by solid gray lines. The dashed vertical line at $\hat{x}_3^+=5$ separates viscous sublayer and buffer layer, the dashed line at $\hat{x}_3^+=30$
separates the buffer and logarithmic layer. The linear law $\hat{u}_1^+=\hat{x}_3^+$ and the von K\'{a}rm\'{a}n law $\hat{u}_1^+=k^{-1}\log \hat{x}_3^++B$
with the constant $k=0.4$ and $B=5.5$ are indicated as solid red lines.  Bottom: Replot of the data in a double linear plot, but with the x-axis as  $\hat{x}_3$. The dashed vertical 
line indicates now the thermal boundary layer thickness which is given by $\delta_T=1/(2 Nu)$.}
\label{fig7}
\end{figure}

Based on the friction velocity (\ref{flux4}), we define an inner viscous scale by 
\begin{equation}
\label{flux4a}
\ell_{\tau}= \sqrt{\frac{Pr}{Ra}} \, u_{\tau}^{-1}\,.
\end{equation}
Thus $\hat{u}_i^+=\hat{u}_i/u_{\tau}$ and $\hat{x}_i^+=\hat{x}_i/\ell_{\tau}$. Note that for Figs. \ref{fig7} and  \ref{fig9}, the instantaneous profiles 
for $\hat{u}_i^+$ versus $\hat{x}_i^+$ are first found by dividing  $\hat{u}_i$ by the instantaneous $u_{\tau}$ and $\hat{x}_i$ by the instantaneous 
$\ell_{\tau}$. Then these profiles are time averaged (see \cite{Schumacher2016} for more details). We can now define a friction Reynolds number as $Re_{\tau}=\tilde{u}_{\tau} \tilde{\delta}_{\ast}/\tilde\nu$. In dimensionless form this translates to
\begin{equation}
\label{Re1}
Re_{\tau}=u_{\tau}\delta_{\ast}\,\sqrt{\frac{Ra}{Pr}}\,.
\end{equation}
\begin{figure}
\centering
\includegraphics[width=0.75\textwidth]{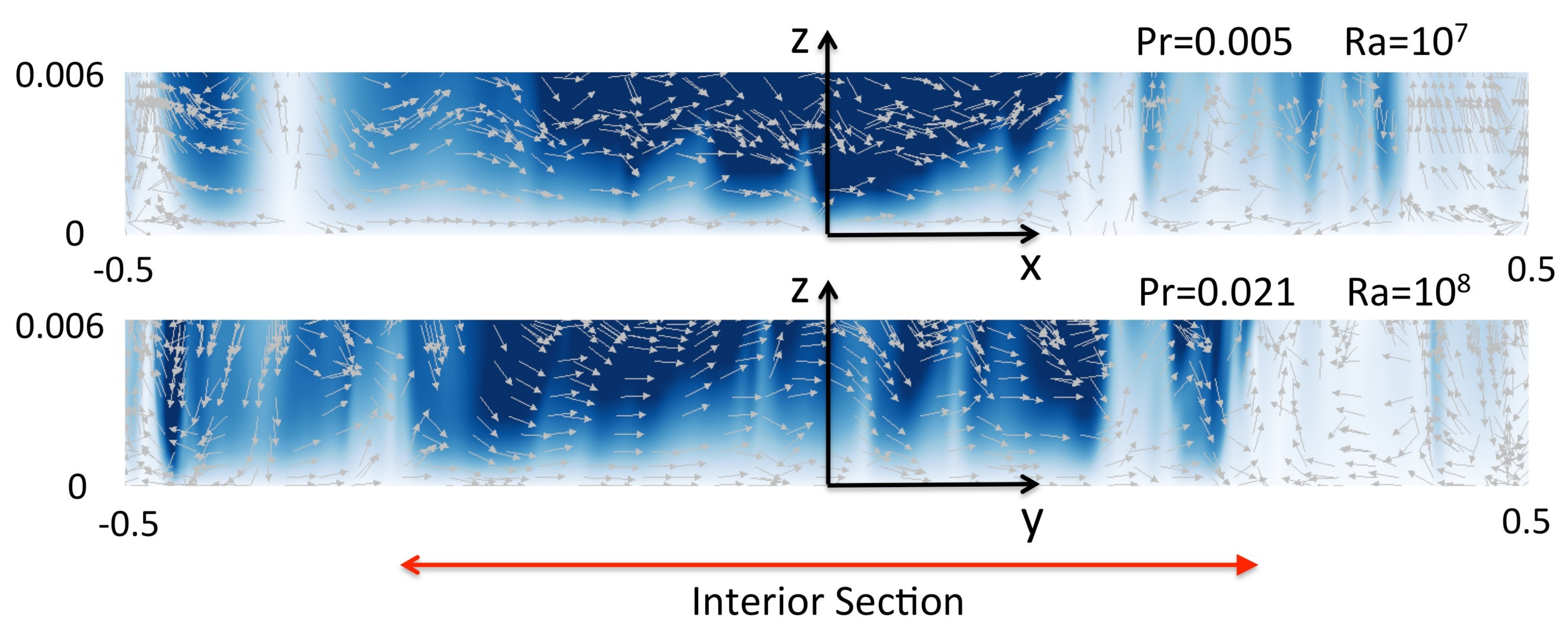}
\caption{Velocity field snapshots inside the thermal boundary layer for runs 14 (bottom) and 19 (top). 
Vertical cuts are taken along the $x$-axis in the top panel and along the $y$-axis in the bottom panel. All arrows are normalized to one 
such that only the flow direction is indicated. The background color stands for the three-dimensional velocity magnitude $u=(u_i^2)^{1/2}$. Colors range from $u=0$ (white) to
$u=0.5$ (deep blue). Note the real aspect ratio of this cut is 167:1 and that the viscous boundary layer thickness is about 1/3 of the displayed height in both cases.
We also indicate the interior section $b$ that is taken for most of the present boundary layer analysis.}
\label{BLsnap}
\end{figure}
The relevant length scale used here is the $x_3$ coordinate of the maximum of the time-averaged velocity profile which is denoted as $\delta_{\ast}$.
Thus $Re_{\tau}$ differs from the typical definition in a turbulent channel flow (CF) where the half width of the channel is taken as the characteristic scale.
Fig. \ref{fig7} shows the mean streamwise velocity profile obtained for the interior plate sections at the top and bottom. The linear relation and the von K\'{a}rm\'{a}n
law are also indicated together with a CF Benchmark case for a turbulent boundary layer at $Re_{\tau}=180$ \cite{Alamo2003}. The RBC flow results in a
friction Reynolds number of $Re_{\tau}=48$. The bottom figure replots data on doubly linear axes. For all RBC data sets, the shape of the mean profile 
is qualitatively the same. In contrast to a pressure-driven CF, the profile passes through a maximum at $\hat{x}_3=\delta_{\ast}$ and decreases to a bulk 
plateau afterwards. The behavior is thus closer to a turbulent plane wall jet as discussed for example by George et al. \cite{George2000}. 

This calls for a closer inspection. Figure \ref{BLsnap} shows typical vertical cuts through the near-wall flow region where we display the velocity field  in a plane along the $y$-axis (run 19 at the top and run 14 at the bottom of the figure). In both instances the flow direction is from left to right in the interior
section which is indicated by the red double-headed arrow. The thickness of the BL varies significantly along the downstream direction, and this  is caused 
by detaching plumes and downwelling from the vigorous bulk turbulence. These processes will prohibit any self-similar downstream evolution as it is known 
from evolving turbulent plane wall jets \cite{George2000}. We have tried without success to collapse the data after re-scaling by $\delta_{\ast}$ and the mean
streamwise velocity maximum.         
\begin{figure}
\centering
\includegraphics[width=0.45\textwidth]{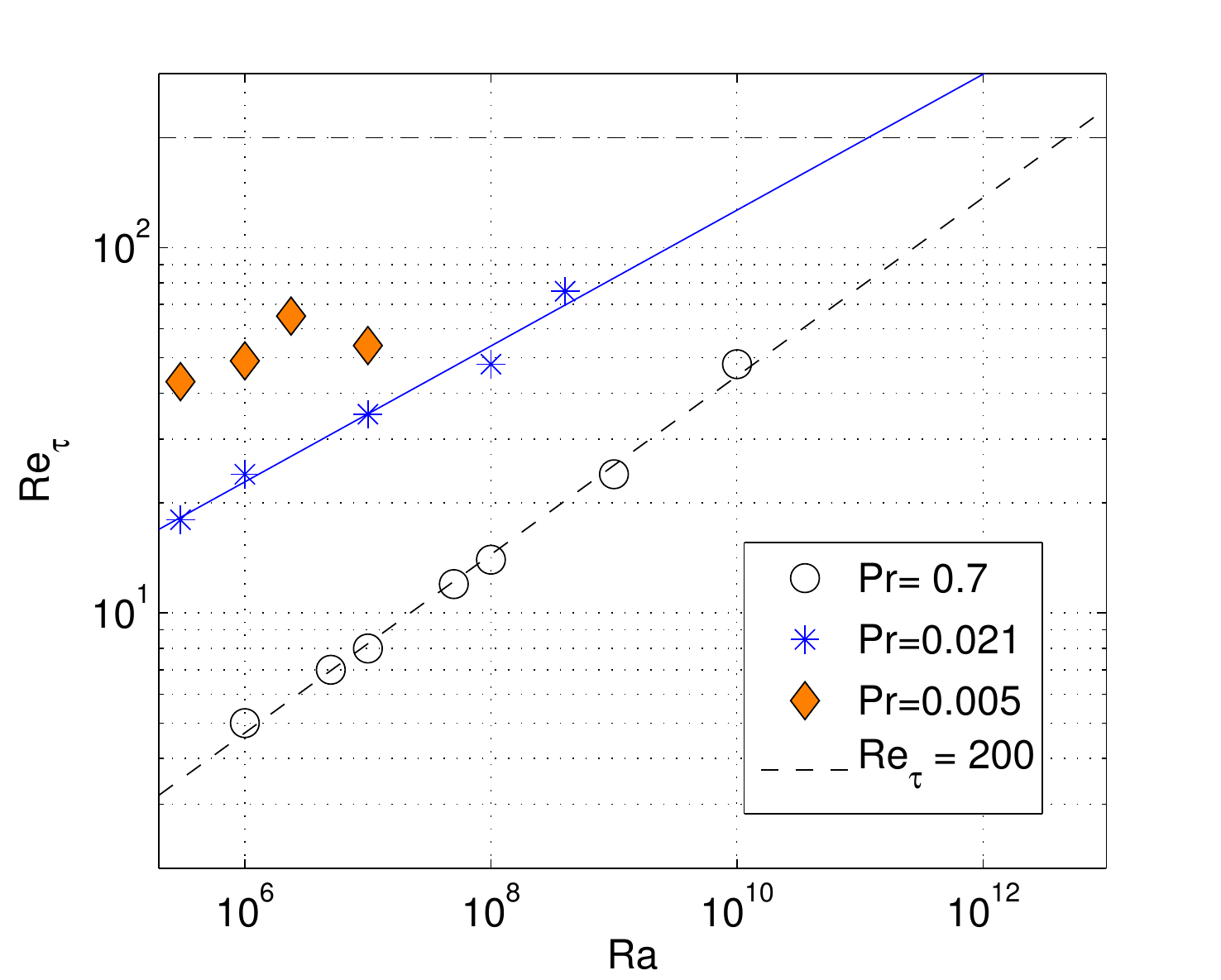}
\caption{Friction Reynolds numbers versus Rayleigh number for three different Prandtl numbers. The horizontal dashed line marks 
$Re_{\tau}=200$.}
\label{fig8}
\end{figure}

Fig. \ref{fig8} shows the resulting friction Reynolds numbers versus Rayleigh number. The series for $Pr=0.021$ and $Pr=0.7$ follow a power law scaling.
By means of an extrapolation one can now determine the range of critical Rayleigh numbers, $Ra_c^{(I)}$ for which $Re_{\tau}\approx 200$, i.e., the friction 
Reynolds number for which the boundary layers in a channel flow are fully turbulent \cite{Elsnab2011,Kim1987} and for which we would expect a transition
into the ultimate convection regime in the present setup. 
This friction Reynolds number is reached in the series for convection in air for an intersection point of 
$Ra^{(I)}_c(Pr=0.7)=5\times 10^{12}$, and a range of 
$10^{12}\le Ra_c \le 2\times 10^{13}$ when the error bars of the power law fit to $Re_{\tau}(Ra)$ are incorporated. The upper boundary of this 
uncertainty range is 
consistent with the convection experiments of He {\it et al.} \cite{He2012,Ahlers2017}. 
In case of mercury, one obtains an intersection point of $Ra_c^{(I)}(Pr=0.021)=10^{11}$ and a range of  $5\times 10^{9}\le Ra_c \le 
4\times 10^{12}$. Note that runs 14 and 15 have been run longer for the present analysis which caused a slightly 
modified range in comparison to Ref. \cite{Schumacher2016}.  The case of liquid sodium is more uncertain. If one includes only the first three data points in the 
least squares fit, an intersection point of $Ra_c^{(I)}(Pr=0.005)=10^{9}$  and a range of $10^{7}\le Ra_c \le 5\times 10^{11}$ results. 
The scatter becomes even larger when the fourth data point of run 19 is incorporated. The extrapolations show that the crossover range is shifted towards smaller Rayleigh numbers as the Prandtl number decreases.   
\begin{figure}
\centering
\vspace{0.5cm}
\includegraphics[width=0.32\textwidth]{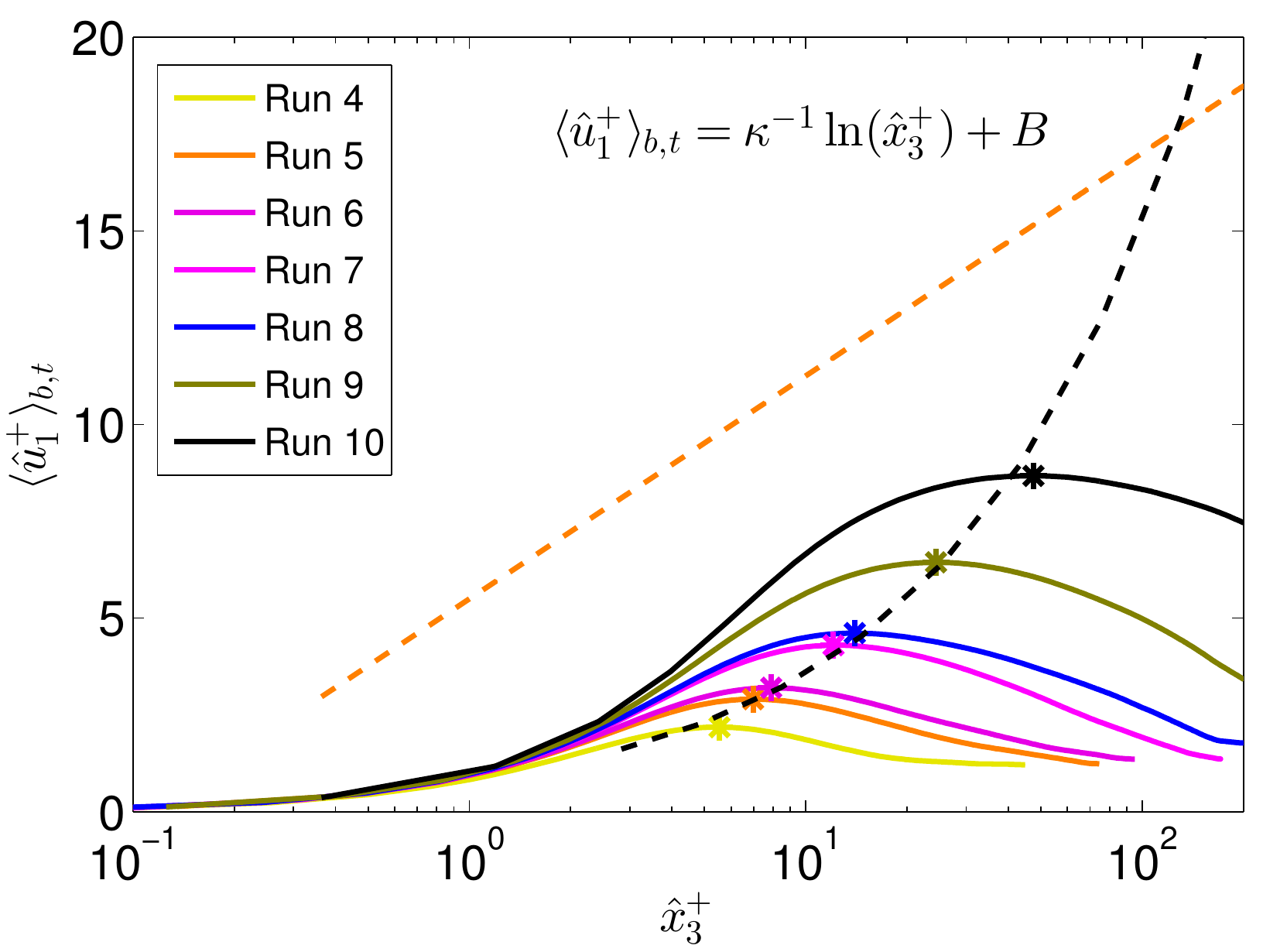}
\includegraphics[width=0.32\textwidth]{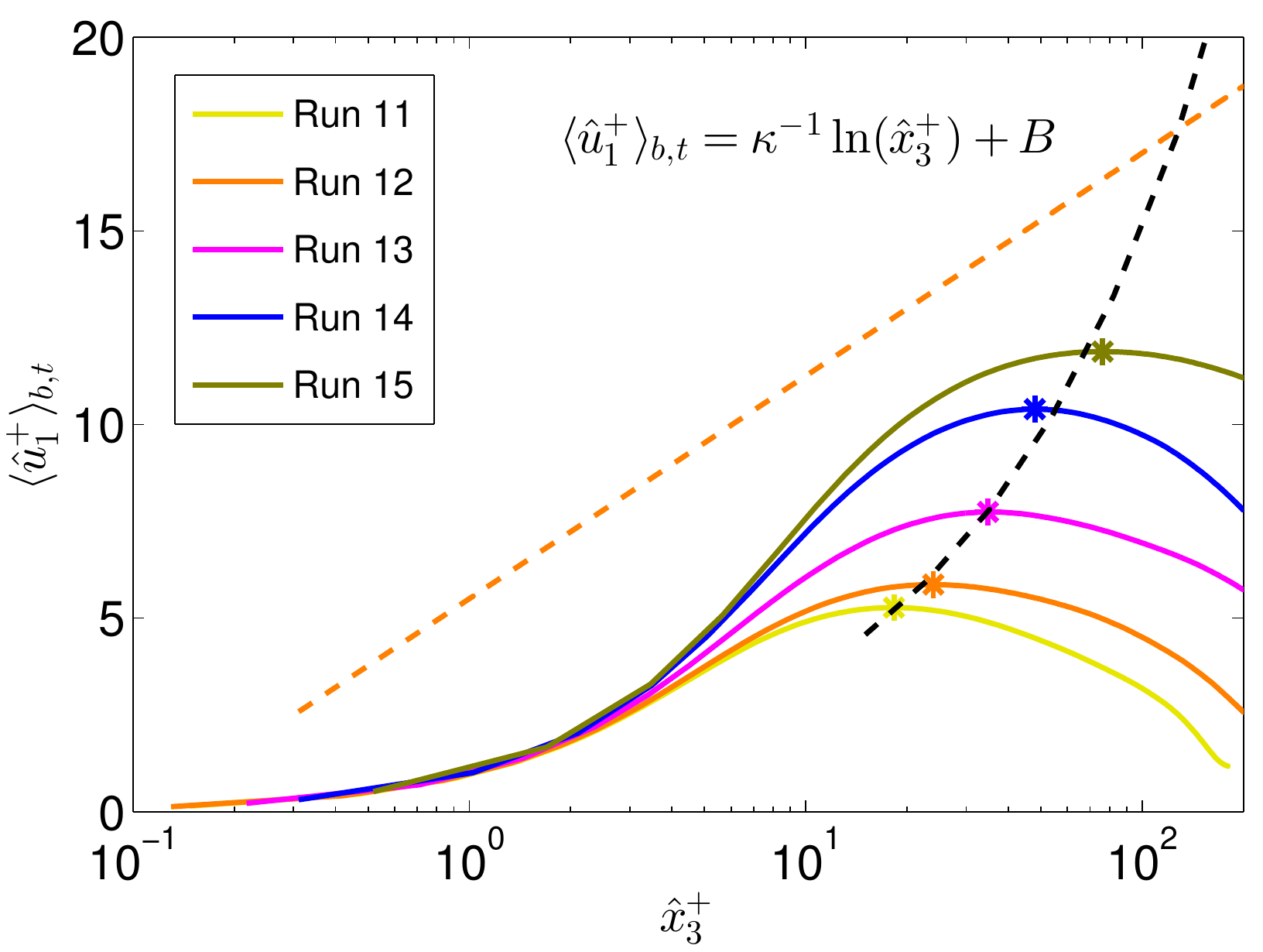}
\includegraphics[width=0.325\textwidth]{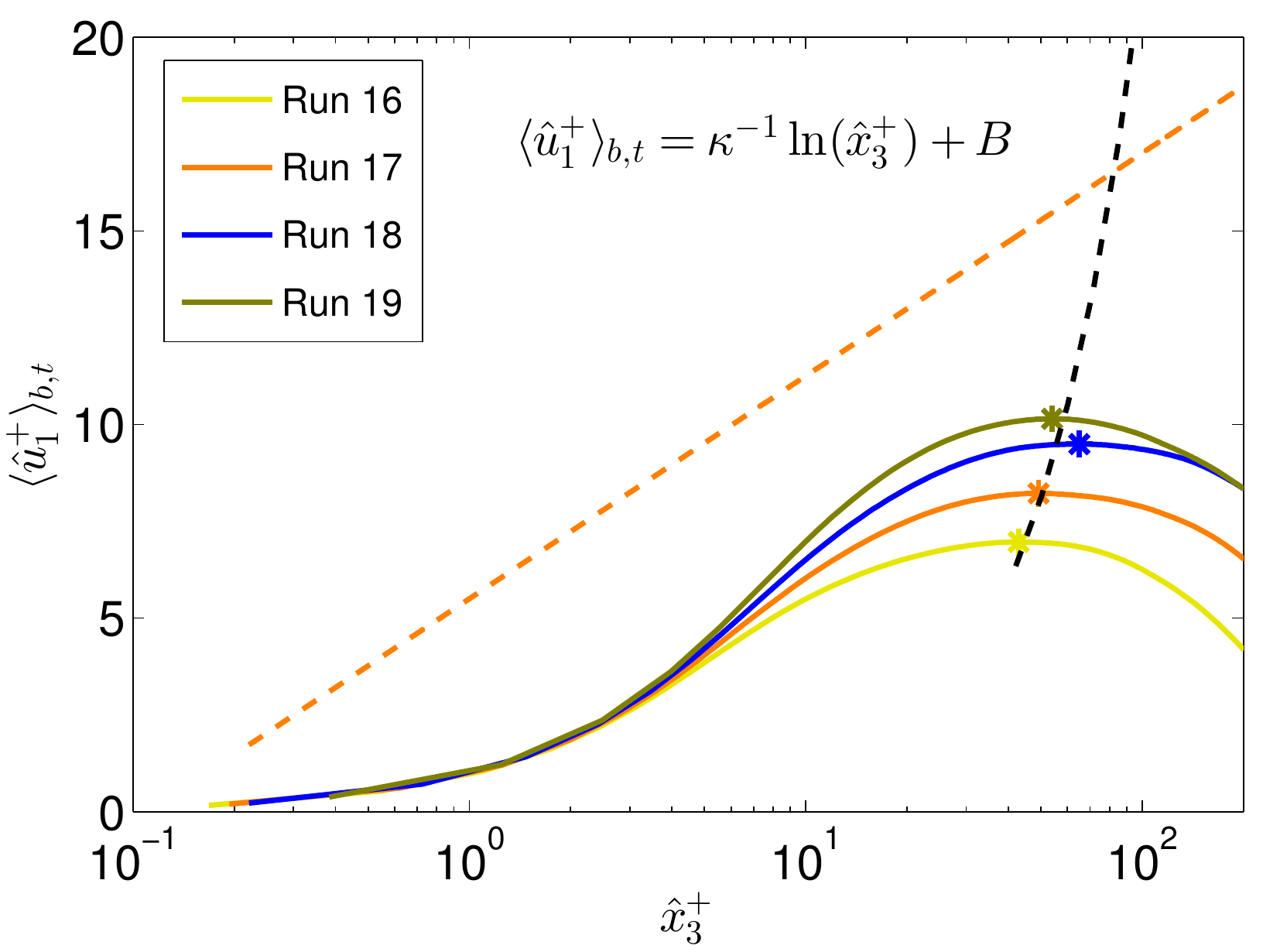}
\caption{Mean streamwise velocity profiles for different Prandtl and Rayleigh numbers (see Tab. \ref{Tab1}). The maxima 
$(\delta_{\ast}/\ell_{\tau}, \hat{u}_3^+(\delta_{\ast}/\ell_{\tau}))$ are indicated as stars. A power law fit to the successive maxima 
in each series is used to predict the intersection point of the logarithmic profiles with the von K\'{a}rm\'{a}n law. This intersection 
point is at  $\hat{u}_1^+ = 134, 129, 83$ for $Pr=0.7, 0.021, 0.005$, respectively. In all cases, the 
profiles taken from the top and bottom plates are included.}
\label{fig9}
\end{figure}
\begin{figure}
\centering
\vspace{0.5cm}
\includegraphics[width=0.45\textwidth]{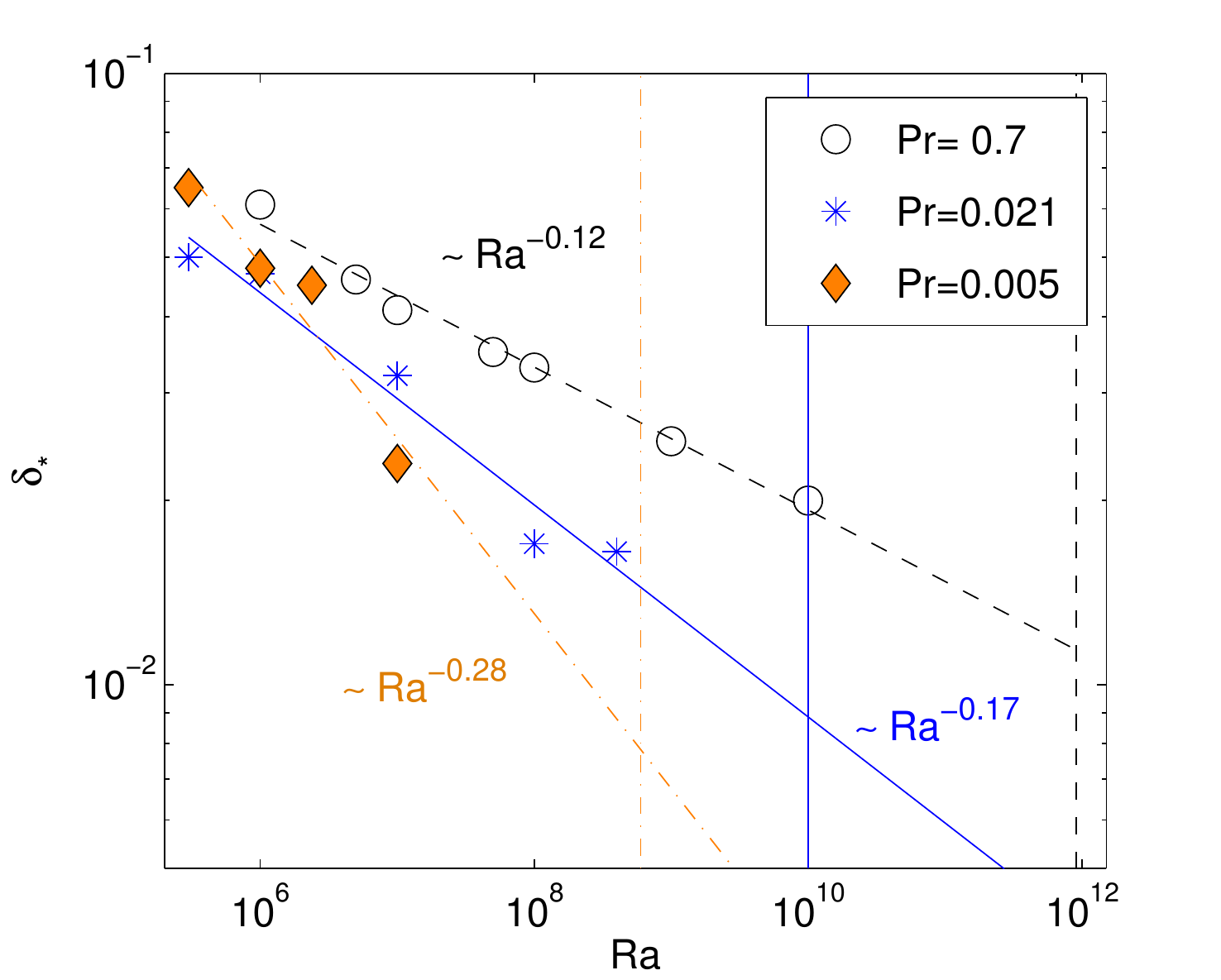}
\caption{Distance of the mean streamwise velocity maximum from the wall, $\delta_{\ast}$, as a function of Rayleigh and Prandtl number. Line styles
agree with those from Fig. \ref{fig8}. The vertical lines (same linestyle as data) correspond to the critical Rayleigh numbers from Tab. II, method II.}
\label{fig10}
\end{figure}
  
\subsection{Method II: Extrapolation of velocity profile maximum}
An alternative method is based on the maxima at $x_3=\delta_{\ast}$ of the streamwise velocity profiles. In Fig. \ref{fig9} we display all mean streamwise velocity 
profiles and indicate the maxima of the velocity profiles for each data set. A curve fit  through the data is also drawn that intersects the law of the wall at a certain 
point. The curve fit is obtained as follows: We fit a power law to the maxima position, ${\delta_{\ast}/\ell_{\tau}}=C_1\times Ra^{c_1}$. The procedure is repeated for 
$\hat{u}_3^+({\delta_{\ast}/ \ell_{\tau}}) =C_2\times Ra^{c_2}$. The result is the parametric dashed curve that connects the maxima and can be extrapolated to higher Rayleigh numbers such that the intersection 
point with the von K\'{a}rm\'{a}n law and thus the corresponding Rayleigh number can be determined. 
  With this procedure we obtain $Ra^{(II)}_c(Pr=0.005)=6\times 10^8$ and an uncertainty range of $3\times 10^6\le Ra_c\le 10^{12}$ for convection in liquid sodium. In case of mercury,  $Ra^{(II)}_c(Pr=0.021)=10^{10}$ and an uncertainty range of $5\times 10^9\le Ra_c\le 5\times 10^{10}$  follows. For convection
in air one obtains $Ra^{(II)}_c(Pr=0.7)=9\times 10^{11}$ and an uncertainty range of $1\times 10^{11}\le Ra_c\le 4\times 10^{12}$.  This method underestimates  $Ra^{(II)}_c$  as is assumed that for this intersection point, the boundary layer is   fully turbulent. However, it may be that the profile needs to overlap with the von K\'{a}rm\'{a}n law for a more substantial region for the boundary layer to be considered fully turbulent. It is seen, however,  that the transition 
ranges of method II overlap significantly with those from method I. Again the uncertainty of the series with lowest Prandtl number is largest. 
The error bars of the power law fits results in the ranges for $Ra_c$. Nevertheless, method II is an appropriate 
complementary way to obtain transition ranges based on large-scale quantities such as mean profiles. For completeness 
we plot $\delta_{\ast}(Ra,Pr)$ in Fig. \ref{fig10}. While the data for $Pr=0.7$ and $0.021$  follow a consistent decrease in magnitude and increase 
in magnitude of scaling exponent, the data for $Pr=0.005$ appears to be more of an outlier. More data at higher $Ra$ and lower $Pr$ would enable us to better understand this behavior.

\subsection{Method III: Shape Factors}

In refs. \cite{Scheel2012, WSPS2015}, we obtained the temperature shape factors for $Pr=0.7$ and $Pr=0.021$. We refine this method and include the results for $Pr=0.005$ 
here. Following Schlichting \cite{Schlichting2000}, we can define a displacement thickness as 
\begin{equation}
\label{displv}
d_u=\int_0^{\delta_{u_h}^*}\left(1-{\langle u_h(x_3) \rangle_{b,t}\over \left[\langle u_h\rangle_{b,t}\right]_{\rm max}}\right) dx_3
\end{equation}
where $\delta_{u_h}^*$ is the $x_3$-location of $ [\langle u_h(x_3)\rangle_{b,t}]_{\rm max}$. The variable $u_h = \sqrt{u_1^2+u_2^2}$ is the rms horizontal velocity.
Likewise we can define a displacement thickness for temperature as
\begin{equation}
\label{displt}
d_T=\int_0^{\infty}\left(1-{\langle T(x_3) \rangle_{b,t}\over \left[\langle T \rangle_{b,t}\right]_{\rm mean}}\right) dx_3
\end{equation}
where $[\langle T \rangle_{b,t}]_{\rm mean}$ is the mean temperature in the bulk. Since the mean temperature profiles are typically not perfectly constant in the bulk (i.e. there tends to be a temperature gradient, which changes with Rayleigh and Prandtl number), we only integrate in practice out to 2.5 times the thermal boundary layer thickness, found here by the slope method. This is the same method used in \cite{Scheel2012}. In Fig. 12 these displacement thicknesses are plotted as a function of Rayleigh and Prandtl numbers. In the left panel, one sees that the thermal displacement thickness increases in magnitude as Prandtl number decreases, and that the scaling exponents systematically decrease, also as Prandtl number decreases. The opposite is found for the velocity displacement thickness: the magnitude is smaller and the exponent is larger as Prandtl number decreases.
This trend is similar to what was found for the local method of finding boundary layer thicknesses as presented in Scheel and Schumacher \cite{Scheel2016}, although the exact exponent values differ.

For a laminar boundary layer \cite{Wilcox2007}, it is expected that the  displacement thickness falls off with Rayleigh number with an exponent of 1/4, but for a turbulent boundary layer this exponent decreases to 1/10. This displacement thickness analysis suggests that all of our boundary layers are closer to a laminar BL, with possibly a slight increase in intermittency as Prandtl number increases for the velocity boundary layer and the opposite trend for the thermal boundary layer.
\begin{figure}
\centering
\vspace{0.5cm}
\includegraphics[width=0.45\textwidth]{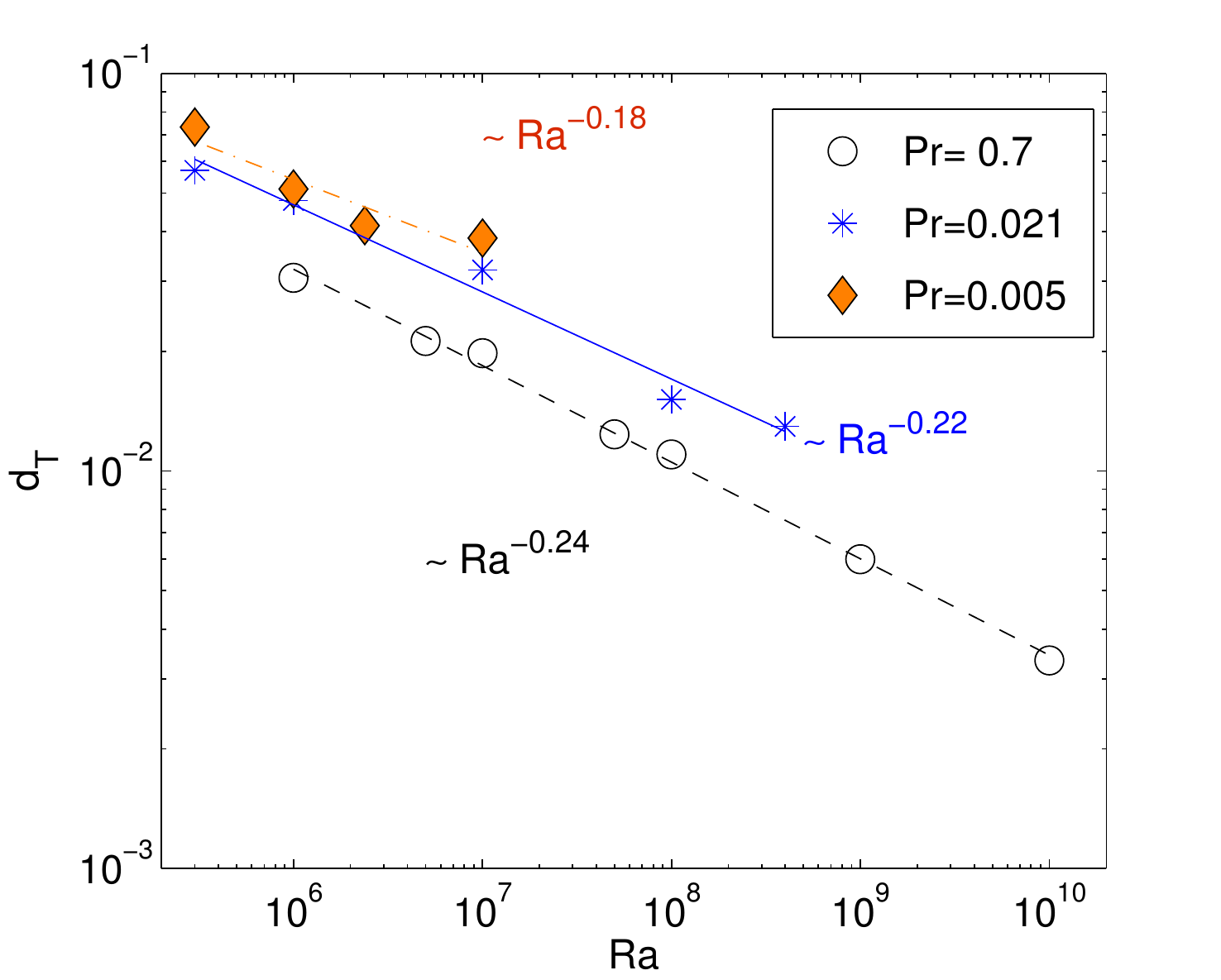}
\includegraphics[width=0.45\textwidth]{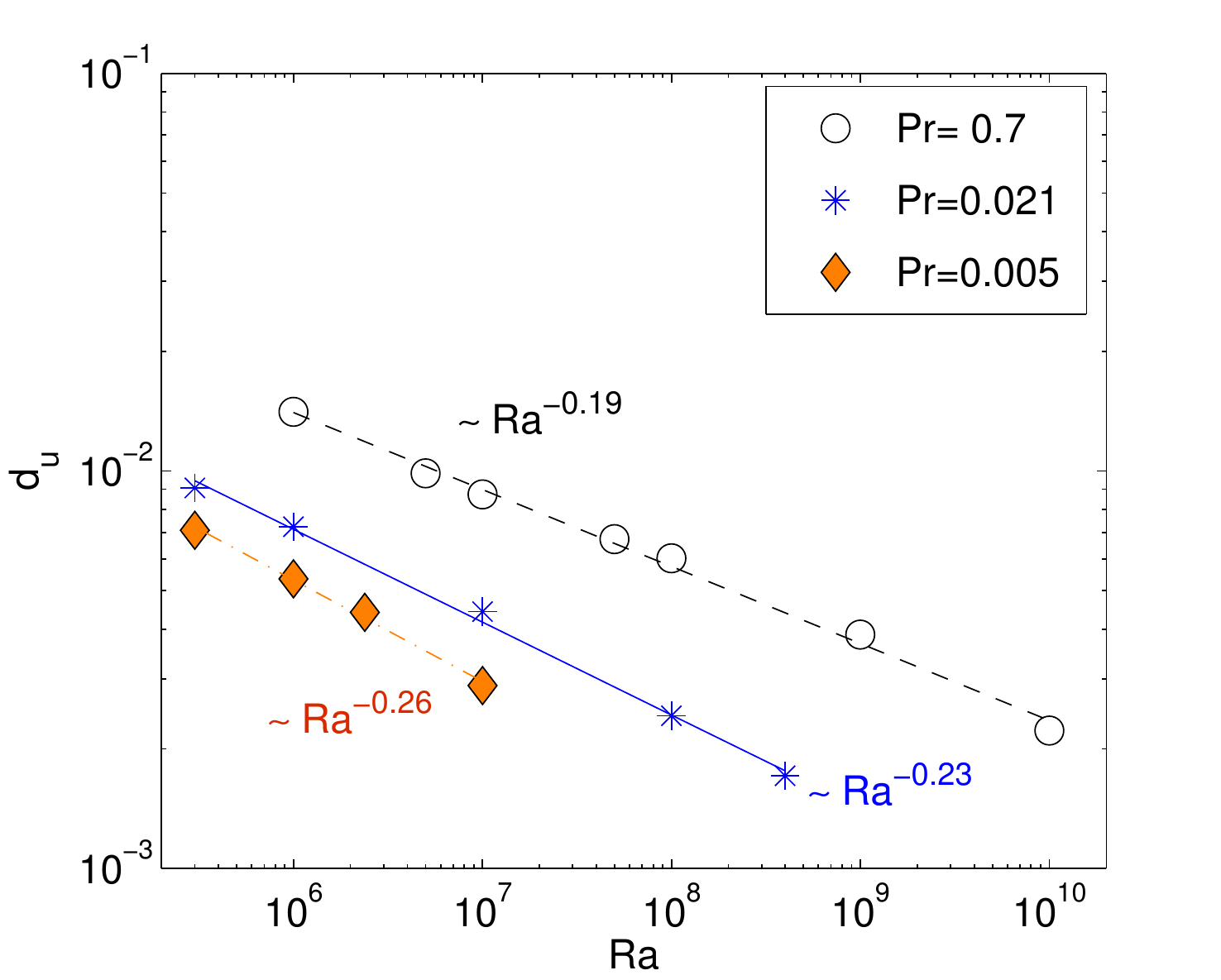}
\label{displace}
\caption{Displacement thicknesses $d_T$ and $d_u$ (found from Eqns. (\ref{displt}) and (\ref{displv}), respectively) as a function of Rayleigh and Prandtl number. }
\end{figure}

A momentum thickness can also be defined for velocity
\begin{equation}
\label{momv}
m_u=\int_0^{\delta_{u_h}^*}\left(1-{\langle u_h(x_3) \rangle_{b,t}\over \left[\langle u_h\rangle_{b,t}\right]_{\rm max}}\right)\left({\langle u_h(x_3) \rangle_{b,t}\over \left[\langle u_h\rangle_{b,t}\right]_{\rm max}}\right) dx_3
\end{equation}
and temperature
\begin{equation}
\label{momt}
m_T=\int_0^{\infty}\left(1-{\langle T(x_3) \rangle_{b,t}\over \left[\langle T \rangle_{b,t}\right]_{\rm mean}}\right) \left({\langle T(x_3) \rangle_{b,t}\over \left[\langle T \rangle_{b,t}\right]_{\rm mean}}\right) dx_3
\end{equation}
From these quantities, we can define the shape factors $H$ as
\begin{equation}
\label{shapefac}
H_T={d_T\over m_T} \quad\quad\mbox{and}\quad\quad H_u={d_u\over m_u}\,.
\end{equation}

\begin{figure}
\centering
\vspace{0.5cm}
\includegraphics[width=0.45\textwidth]{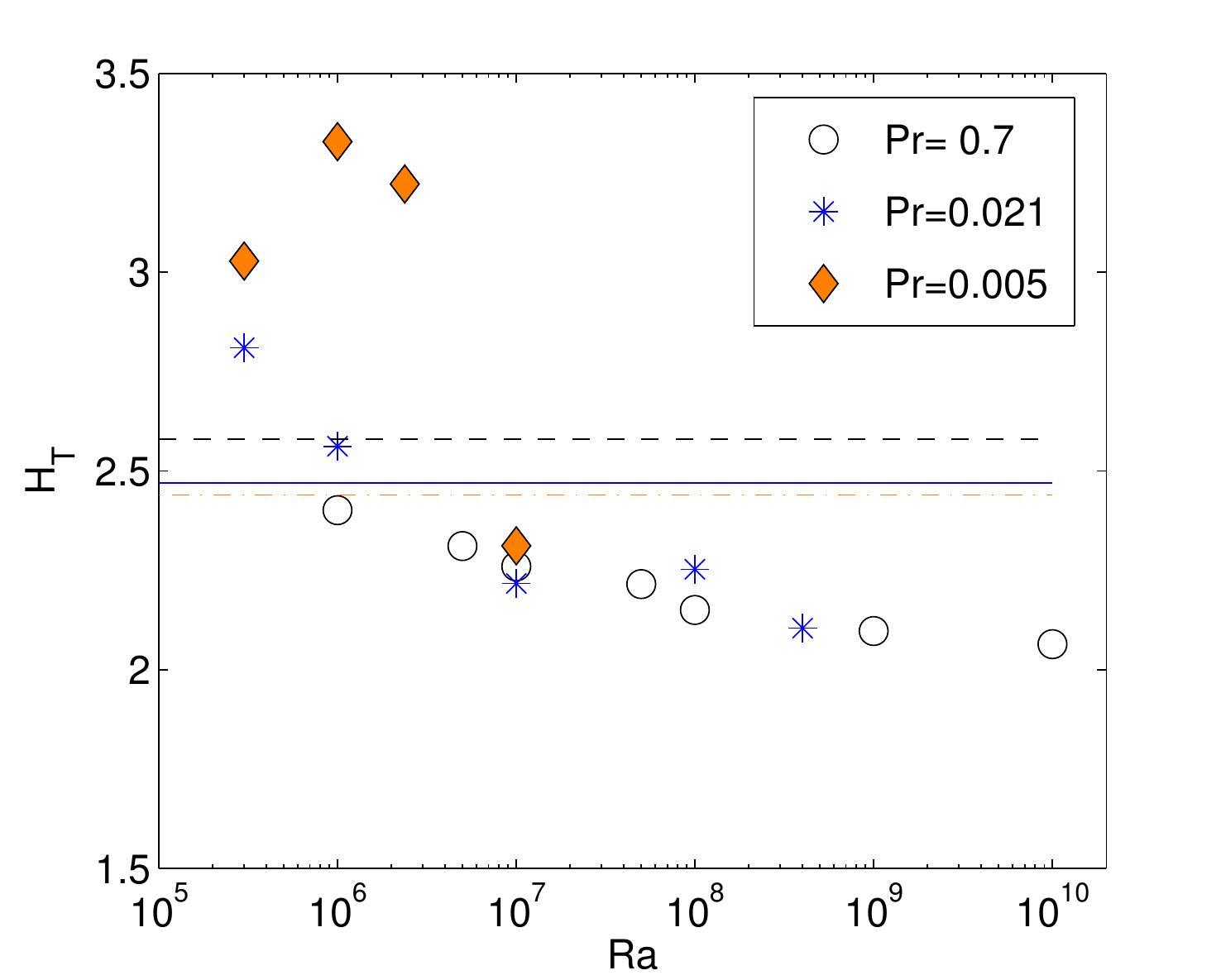}
\includegraphics[width=0.45\textwidth]{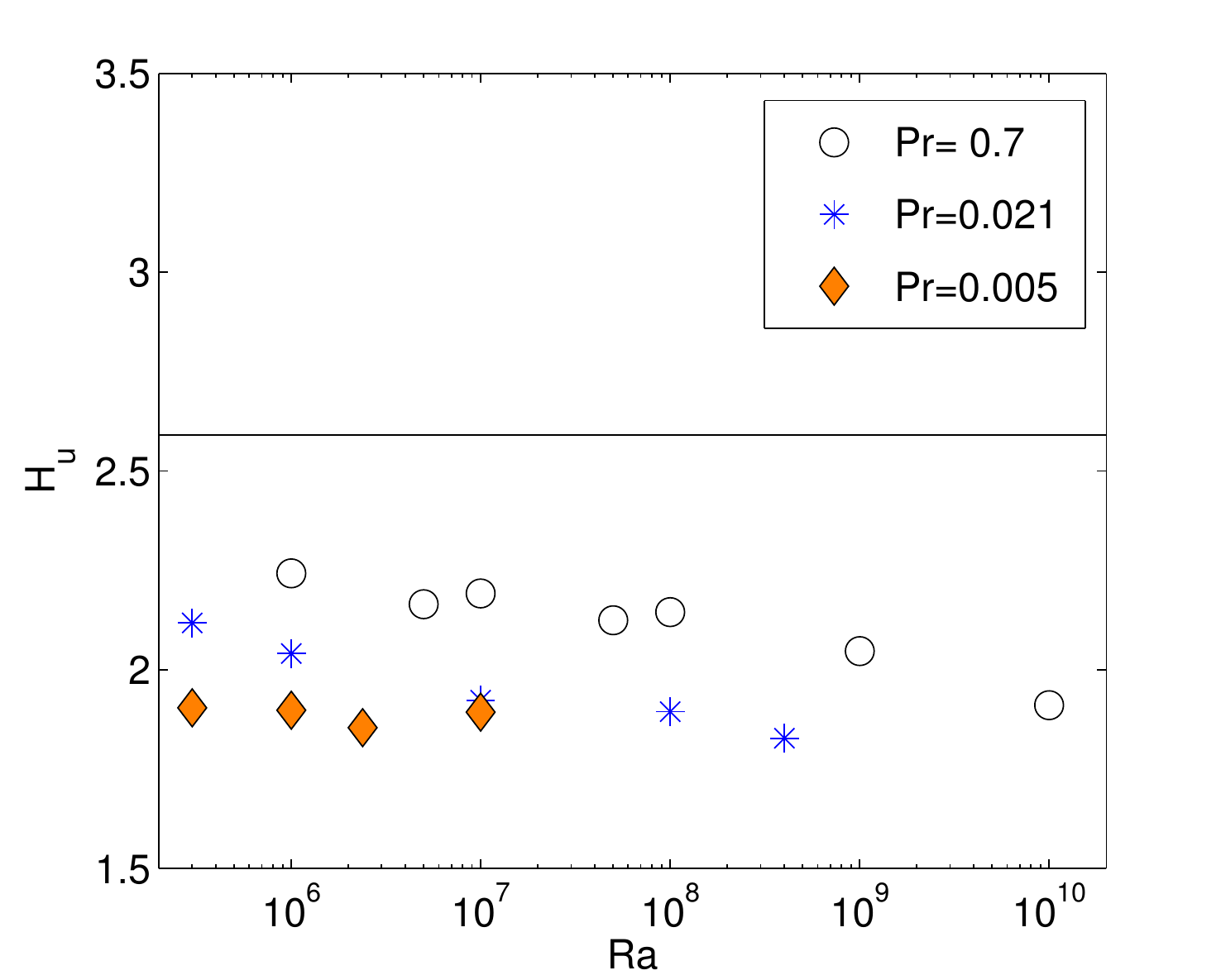}
\caption{Shape factors, $H_T$ (left) and $H_u$ (right) as defined in Eqn. (\ref{shapefac}) as a function of Rayleigh and Prandtl numbers. All horizontal lines are the shape factors for the Prandtl-Blasius case. For temperature, the shape factor $H_T$ depends on Prandtl number: the dashed line is for $Pr=0.7$, the solid line is for $Pr=0.021$ and the dashed-dotted line is for $Pr=0.005$. For velocity, the shape factor $H_u$ is independent of Prandtl number.}
\label{sf}
\end{figure}

The shape factor  for the laminar Prandtl-Blasius profile is 2.59 for $H_u$, independent of Prandtl number. The computed quantity for  $H_T$ is Prandtl number dependent and is 2.58, 2.47, and 2.44 for $Pr$ = 0.7, 0.021 and 0.005, respectively (see Fig. \ref{sf}). 
A shape factor that is larger than the laminar value tends to indicate an overshoot of the Prandtl-Blasius 
profile and a smaller shape factor indicates an undershoot. The latter is caused by rising (or falling) plumes and resulting recirculations due to 
flow incompressibility. In almost all cases, the shape factors are fairly close, but tend to fall below the Prandtl-Blasius values. 
The results for $H_u$ demonstrate a systematic decrease in shape factor as Prandtl number decreases. The results for $H_T$ show similar shape factor values for $Pr=0.7$ and $Pr=0.021$ as was also presented in \cite{WSPS2015}. However, the temperature shape factor for $Pr=0.005$ is an outlier. And the shape of the 
temperature profile does  look different for $Ra<1\times 10^7$, with a distinct rounded peak at the end of the boundary layer as is evidenced by the larger shape factors. For $Ra=1\times 10^7$, the thermal boundary layer more closely resembles the Prandtl-Blasius profile.

\begin{figure}
\centering
\vspace{0.5cm}
\includegraphics[width=0.5\textwidth]{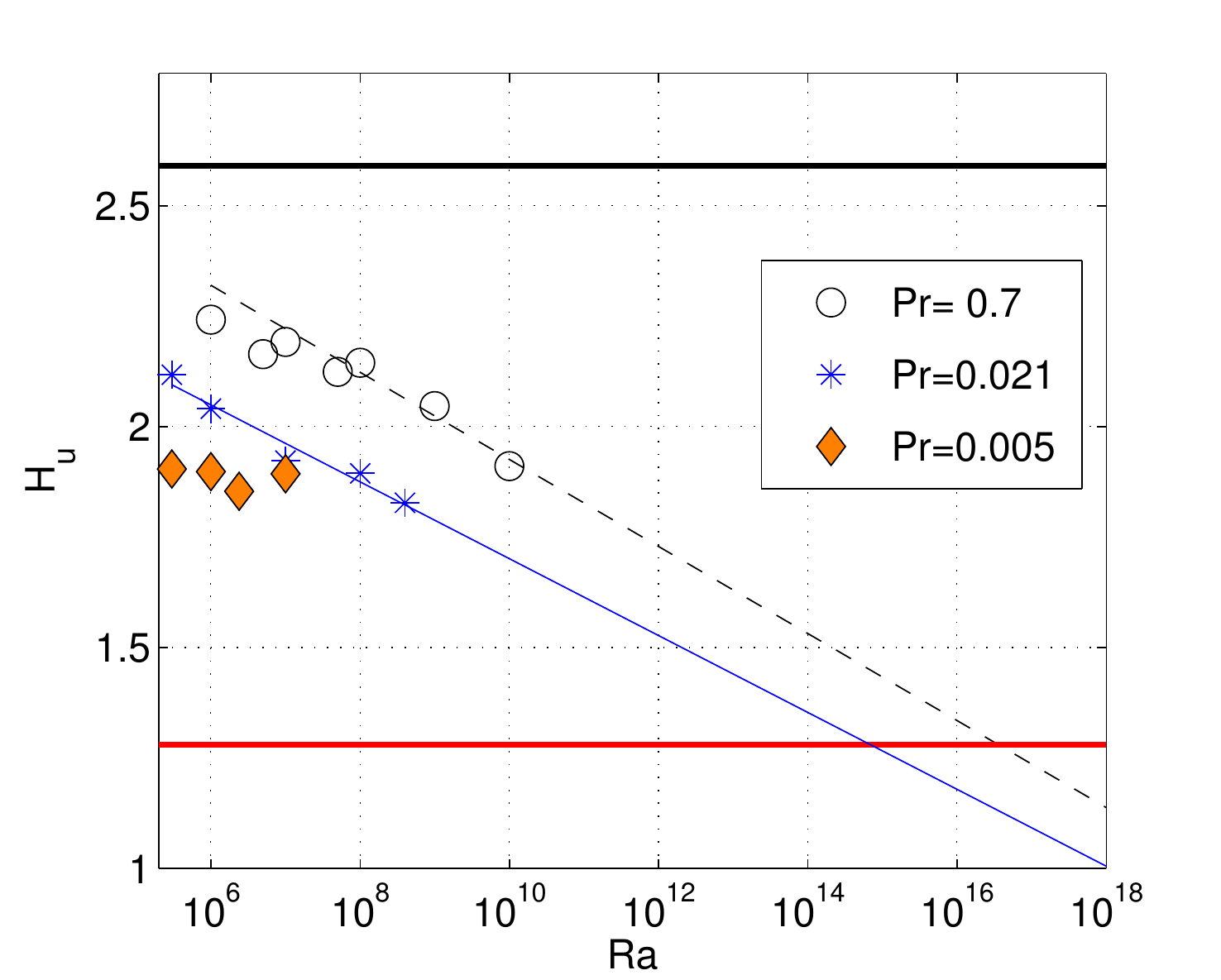}
\caption{Extrapolation of functional dependence of the velocity shape factor to determine $Ra_c^{(III)}$, the critical Rayleigh number for the transition to a turbulent boundary layer. Note for $Pr=0.7$, the fit is only to the last four data points. The horizontal black line at 2.59 is the shape factor for a laminar boundary layer and the horizontal red line at 1.28 is the value of the shape factor for a turbulent boundary layer.}
\label{sfext}
\end{figure}

Finally, we can use our results for the velocity shape factor, $H_u$ to predict a transition to a turbulent boundary layer. The shape factor $H_u$ should be 1.28 for a turbulent boundary layer \cite{Wilcox2007}. In Fig. \ref{sfext} we fit lines to the data and find the intersection of these fits with the horizontal line at $H_u$ = 1.28. This value will be $Ra_c^{(III)}$. For $Pr=0.7$ a fit to the last four data points gives  $Ra_c^{(III)}(Pr=0.7) = 4\times 10^{16}$, for $Pr=0.021$ the $Ra_c^{(III)}(Pr=0.021) = 7\times 10^{14}$. The data for $Pr=0.005$ is very scattered with no clear trend. The resulting uncertainty ranges vary quite widely, when the error associated with the least squares fit is taken into account. These ranges are $6\times 10^{12} <  Ra_c < 1\times 10^{22}$ for $Pr=0.7$ and $2\times 10^{12} <  Ra_c < 1\times 10^{18}$ for $Pr=0.021$. While the predictions for $Ra_c^{(III)}$ are larger  for the other methods,  the values of $Ra_c^{(III)}$ do overlap when the uncertainty is taken into account. It is not surprising that method III overestimates the value for $Ra_c$ since the boundary layer profile would be fully turbulent if $H_u=1.28$, whereas for method II, we extrapolate to where the boundary layer just reaches the log-law, or when it just starts to become turbulent. Furthermore, as summarized in Ref. \cite{Schlatter2012}, the shape factor $H_u$ of turbulent flat plate boundary layers is often found rather at values of 1.4, i.e., larger than 1.28. 
This would additionally decrease the critical Rayleigh numbers in our extrapolation.  The results for all three methods are summarized in Tab. \ref{Tab2}. 
In addition, the results from a fourth method that we discussed in \cite{Scheel2016} are given for completeness. 
This method is based on a shear Reynolds number $Re_{sh}$ which can be determined by means of a root mean square velocity taken at 
the distance from the plate that corresponds to the mean of a locally fluctuating viscous boundary layer thickness. The local boundary 
layer thickness is determined from the vertical derivatives of the horizontal velocity components which correspond to the two components 
of the skin friction vector field. A turbulent fraction is defined as the percentage of BL area for which $Re_{sh}\gtrsim 350$. The transition follows 
for those $Ra$ that would have a turbulent fraction of 100\%.
\begin{table*}
\begin{center}
\begin{tabular}{c  c c c }
\hline\hline
Extrapolation Method & $Pr$ & $Ra_c$ & Uncertainty Range \\ \hline
I & 0.7       & $\quad Ra_c=5\times 10^{12}\quad$  &$1\times 10^{12} < Ra_c < 2\times 10^{13}$   \\
I & 0.021   & $\quad Ra_c=1\times 10^{11}\quad$  &$5 \times 10^{9} < Ra_c < 4\times 10^{12}$    \\
I & 0.005   & $\quad Ra_c=1\times 10^{9}\quad$    &$1 \times 10^{7} < Ra_c < 5\times 10^{11}$      \\ 
II & 0.7      & $\quad Ra_c=9\times 10^{11}\quad$  &$1\times 10^{11} < Ra_c < 4\times 10^{12}$   \\
II & 0.021  & $\quad Ra_c=1\times 10^{10}\quad$  &$5\times 10^{9} < Ra_c < 5\times 10^{10}$    \\
II & 0.005  & $\quad Ra_c=6\times 10^{8}\quad$    &$3\times 10^{6} < Ra_c < 1\times 10^{12}$      \\
III & 0.7     & $\quad Ra_c=4\times 10^{16}\quad$  &$6\times 10^{12} < Ra_c < 1\times 10^{22}$  \\
III & 0.021 & $\quad Ra_c=7\times 10^{14}\quad$  &$2\times 10^{12} < Ra_c < 1\times 10^{18}$ \\ 
IV & 0.7     & $\quad Ra_c=3\times 10^{13}\quad$ &$1\times 10^{13} < Ra_c < 5 \times 10^{13}$  \\
IV & 0.021 & $\quad Ra_c=2\times 10^{13}\quad$ &$1\times 10^{12} < Ra_c < 1\times 10^{14}$   \\ \hline\hline
\end{tabular}  
\caption{A comparison of different methods to predict the range of critical Rayleigh numbers, $Ra_c$, at which the viscous boundary layer becomes turbulent for three different Prandtl numbers. Methods I-III are described in this paper. Method IV was discussed in \cite{Scheel2016}.}
\label{Tab2}
\end{center}
\end{table*}

\section{Temperature field statistics}

\subsection{Mean profiles}
\begin{figure}
\centering
\vspace{0.5cm}
\includegraphics[width=0.5\textwidth]{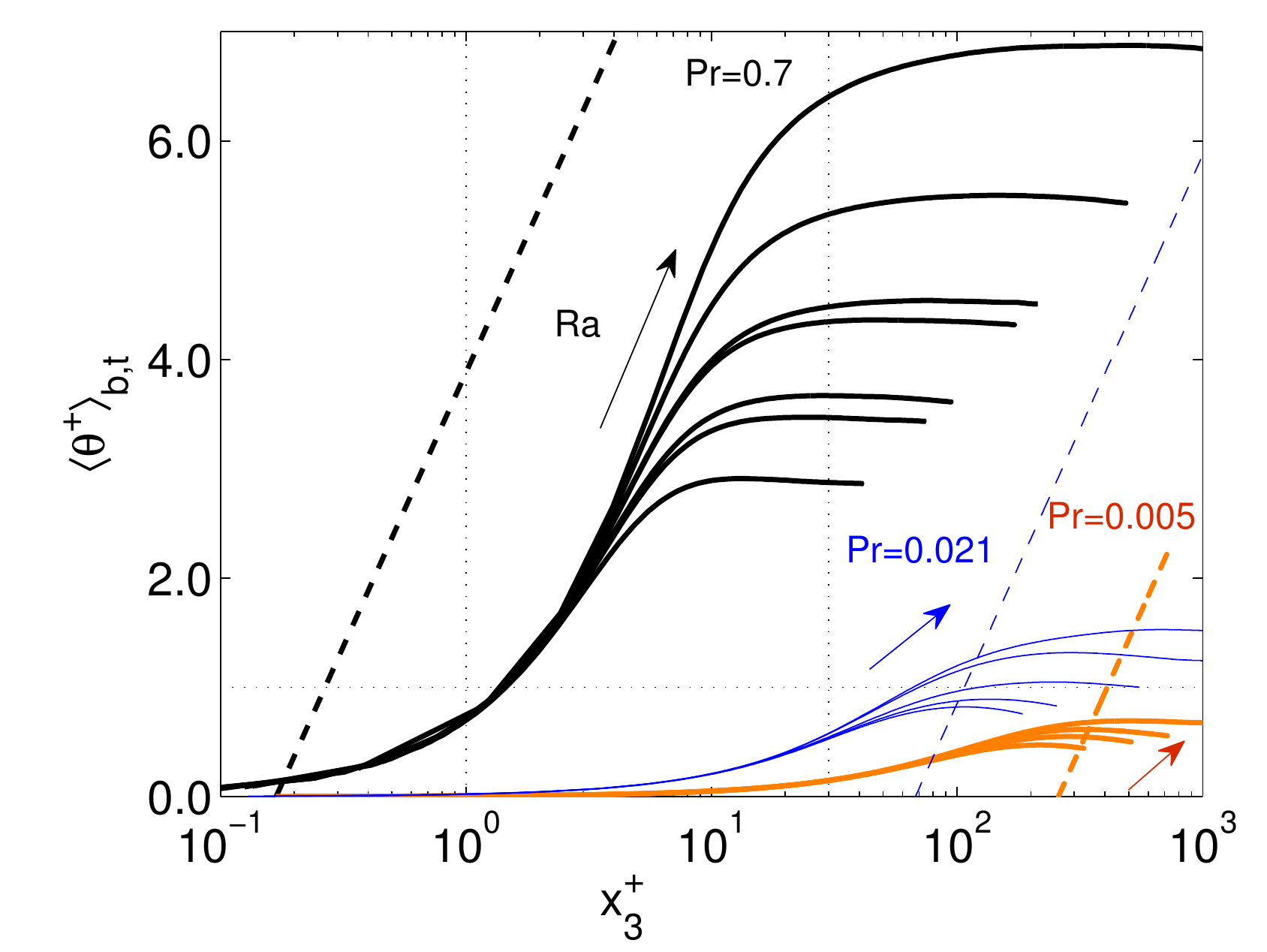}
\caption{Semilogarithmic plot of mean temperature profiles scaled by inner wall units.  These profiles are given for the three indicated Prandtl numbers as a function of Rayleigh number (and increasing Rayleigh number is indicated by an arrow). The dashed lines are  fits to the theoretically predicted law by Yaglom and Kader \cite{Yaglom1979, Kader1981} $\langle\theta^+(x_3^+)\rangle= \alpha \ln x_3^+ + \beta(Pr)$ with $\alpha\approx 2.12\,,\;\;\beta(Pr)=(3.8 Pr^{1/3}-1)^2-1+2.12 \ln Pr$. In all cases, the profiles taken from the top and bottom plates are included.}
\label{tempprof}
\end{figure}

In ref. \cite{Schumacher2016}, we also examined the profiles of the mean temperature inner wall units and we extend this analysis here. We define the friction temperature as
\begin{equation}
\label{temp2}
T_{\tau}=-\Bigg\langle\frac{u_{\tau}^{-1}(t)}{\sqrt{Ra Pr}} \Bigg\langle\frac{\partial T}{\partial x_3}\Bigg\rangle_{b}\Bigg|_{x_3=0}\Bigg\rangle_t\,.
\end{equation}
Thus $\theta^+= T/T_{\tau}$ and we plot these profiles as a function of Rayleigh and Prandtl number in Fig. \ref{tempprof}. Note that for Figs. \ref{tempprof}, the instantaneous profiles for $\theta^+$ versus ${x}_3^+$ are first found by dividing  $\theta=\{T/\Delta T$ for $x_3=1$, $1-{T/\Delta T}$ for $x_3=0\}$ by the instantaneous $T_{\tau}$ and ${x}_3$ by the instantaneous $\ell_{\tau}$.
Then these profiles are time averaged  (see \cite{Schumacher2016} for more details). We see that the profiles follow a similar trend of rising as Rayleigh increases, with a portion of the profile following a logarithmic law. The region of the profile that is logarithmic tends to increase as Rayleigh number increases. The dashed lines are a plot of the theoretically predicted turbulent temperature profile by  Yaglom and Kader \cite{Yaglom1979, Kader1981} for each Prandtl number: $\langle\theta^+(x_3^+)\rangle= \alpha \ln x_3^+ + \beta(Pr)$ with $\alpha\approx 2.12\,,\;\;\beta(Pr)=(3.8 Pr^{1/3}-1)^2-1+2.12 \ln Pr$. While the profiles for $Pr=0.021$ and $Pr=0.005$ have slopes that are smaller than the slope of theoretical log law, the slope for $Pr=0.7$ is quite similar, with an agreement of 10\% for the three highest Rayleigh numbers (although the intercept does not match). Finally one sees that as Prandtl number decreases, the temperature profiles tend to decrease in size in these inner wall units, but they reach their peak at larger $x_3^+$ values. 

\begin{figure}
\centering
\vspace{0.5cm}
\includegraphics[width=0.5\textwidth]{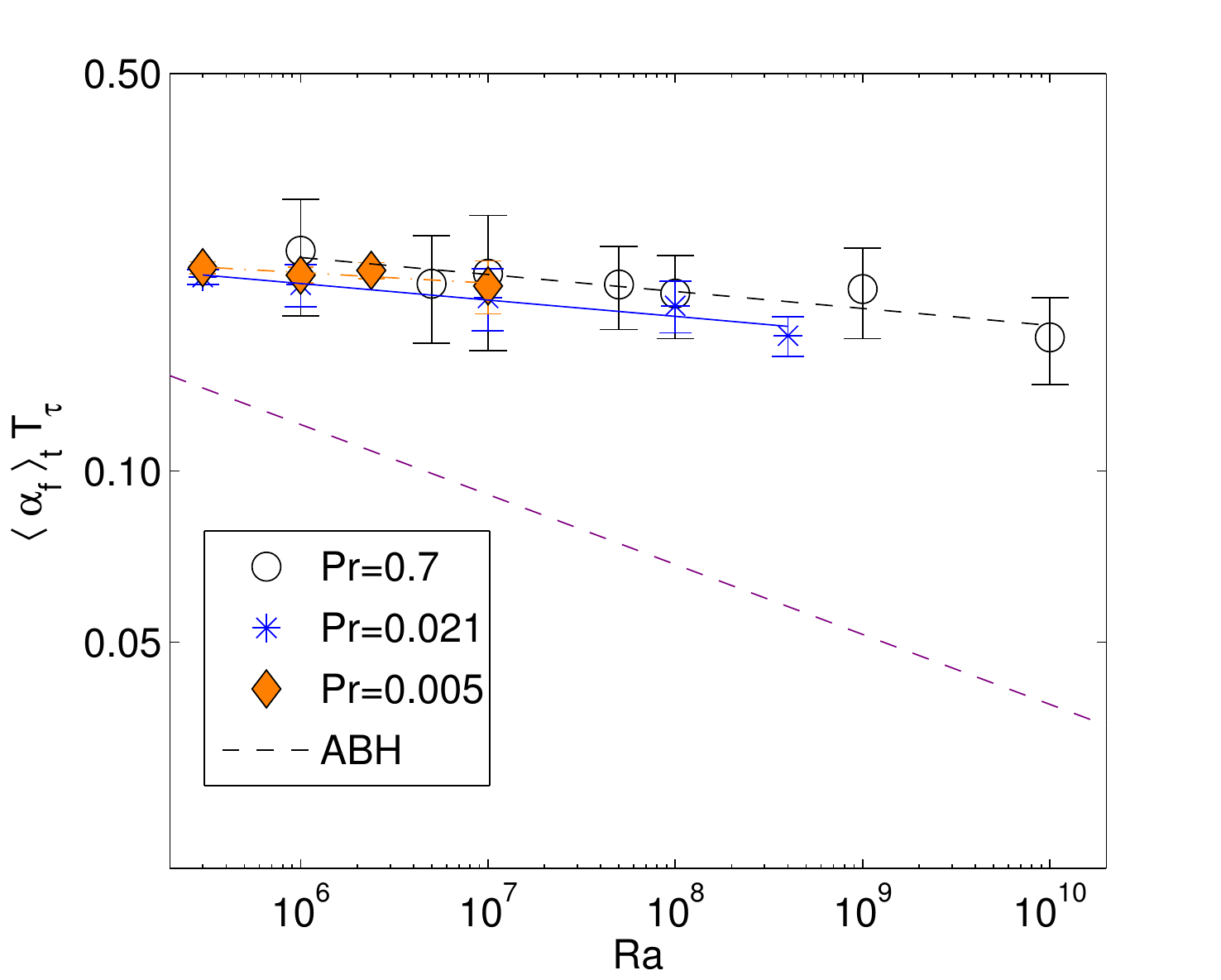}
\caption{Slopes $\langle\alpha_f\rangle_t$ found by fitting a line to the linear portion of Fig. \ref{tempprof} for each Rayleigh and Prandtl number. Then each slope $\langle\alpha_f\rangle_t$ is multiplied by its respective friction temperature $T_{\tau}$. Fits to the data are $\langle\alpha_f\rangle_t T_{\tau}=(1.0 \pm 0.1)Ra^{-0.03 \pm 0.01}$ for $Pr=0.7$ (black dashed line), $\langle\alpha_f\rangle_t T_{\tau}=(0.15 \pm 0.06)Ra^{-0.03 \pm 0.001}$ for $Pr=0.021$ (blue solid line) and  $\langle\alpha_f\rangle_t T_{\tau}=(0.06 \pm 0.03)Ra^{-0.02 \pm 0.01}$ for $Pr=0.005$ (yellow dashed-dotted line). The purple dashed line is the equation by Ahlers et al. (ABH)\cite{Ahlers2014}: $\langle\alpha_f\rangle_t T_{\tau} = (0.66\pm 0.06)Ra^{-0.123\pm 0.01}$.} 
\label{slopeprof}
\end{figure}

The slopes $\langle\alpha_f\rangle_t$ of the linear portion of the semilog plots are found by fitting each instantaneous temperature profile to the equation  $\langle\theta^+(x_3^+)\rangle_{b} = \alpha_f \ln x_3^+ + \beta_f$. The $\langle\alpha_f\rangle_t$ values are then the time averages. The error bars are given by the standard deviation in the $\alpha_f$ values found from the instantaneous profiles. Then these values are converted to units of unscaled $T$ versus $x_3$ by multiplying by the friction temperature $T_{\tau}$. These slopes,  $\langle\alpha_f\rangle_tT_{\tau}$ are plotted in Fig. \ref{slopeprof} for all available Rayleigh and Prandtl numbers. The universality in these slopes for the three Prandtl numbers is quite remarkable. In addition the results from  Ahlers et. al. \cite{Ahlers2014} are also plotted, which were for $Pr=0.8$, $\Gamma = 0.5$  and $5\times 10^{11} < Ra < 5\times 10^{12}$. However, their results were for an aspect ratio of 1/2 and they measured the profiles close to the sidewall of the container, and either of these could account for the discrepancy. In their paper, they also predicted a theoretical scaling exponent between $-0.005$ and $-0.106$, which does agree with our exponent.  In addition, Wei and Ahlers \cite{Wei2014} found, for $Pr=12.3, \Gamma = 1$ and $2\times 10^{10} < Ra < 2\times 10^{11}$, a scaling exponent of $-0.011 \pm 0.025$ (no prefactor given) which is in agreement with our results. More work needs to be done experimentally,  numerically and theoretically to better understand the dependence of  $\langle\alpha_f\rangle_tT_{\tau}$  on Rayleigh and Prandtl number.
\begin{figure}
\centering
\includegraphics[width=0.85\textwidth]{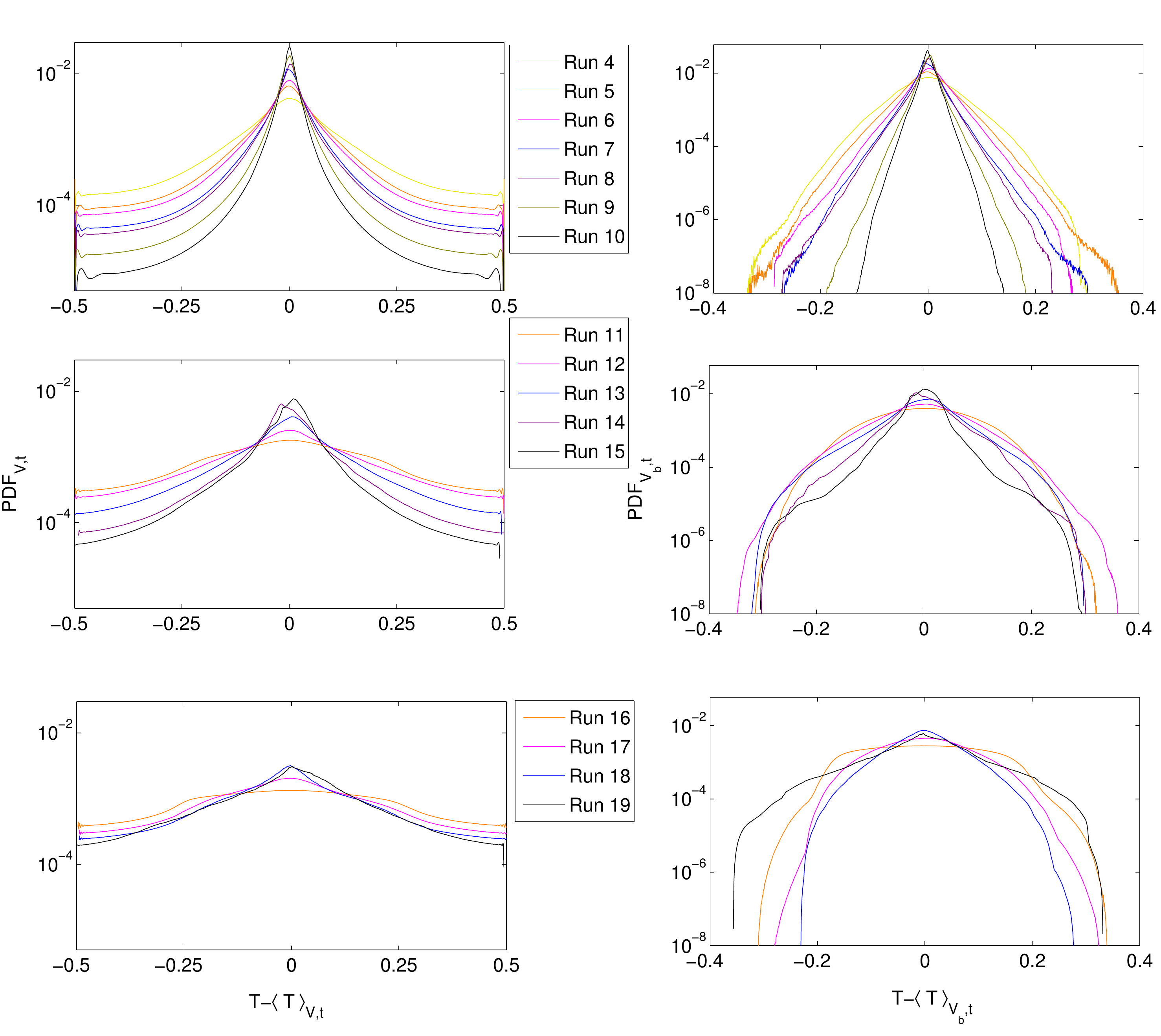}
\caption{Temperature probability density functions (PDFs) for $Pr=0.7$ (top row), $Pr=0.021$, (middle row) and  $Pr=0.005$ (bottom row). The horizontal axis is the temperature minus the global, $\langle T\rangle_{V,t}$, or bulk mean temperature, $\langle T\rangle_{V_b,t}$. The left column (PDF$_{V,t}$)  gives the PDFs taken over the entire container, and the right column (PDF$_{V_b,t}$) gives the PDFs taken over the bulk (see caption for  Fig.~\ref{fig5} for a definition of the bulk).}
\label{fign}
\end{figure}

\subsection{Probability density functions}

Finally we summarize the statistics of the temperature field for the three series of simulation runs. The probability density functions 
(PDFs) of $T(\bm x,t)$ for different Rayleigh and Prandtl numbers are shown in Fig. \ref{fign}. While the panels to the left evaluate the 
distribution of $T$ in the full volume $V$, the data to the right are obtained in the subvolume $V_b$ (see section III). All PDFs are centered 
around their global $\langle T\rangle_{V,t}$ or  bulk mean $\langle T\rangle_{V_b,t}$, respectively.  We find that the temperature PDFs at a 
fixed $Pr$ become systematically more narrow about the mean as the Rayleigh number increases, both, in the bulk and the full volume.
This trend agrees with what is seen in ref.~\cite{Glazier1999} in the bulk  
for the high-Rayleigh number experiments in mercury. However for $Pr=0.005$ and the highest Rayleigh number (run 19), we see that PDF$_{V_b,t}$, 
the distribution taken over the bulk of the container, has widened. The reason for this observation could be that the Reynolds number is large enough to 
generate sharper plume structures of the temperature field.

\begin{figure}
\centering
\vspace{0.5cm}
\includegraphics[width=0.32\textwidth]{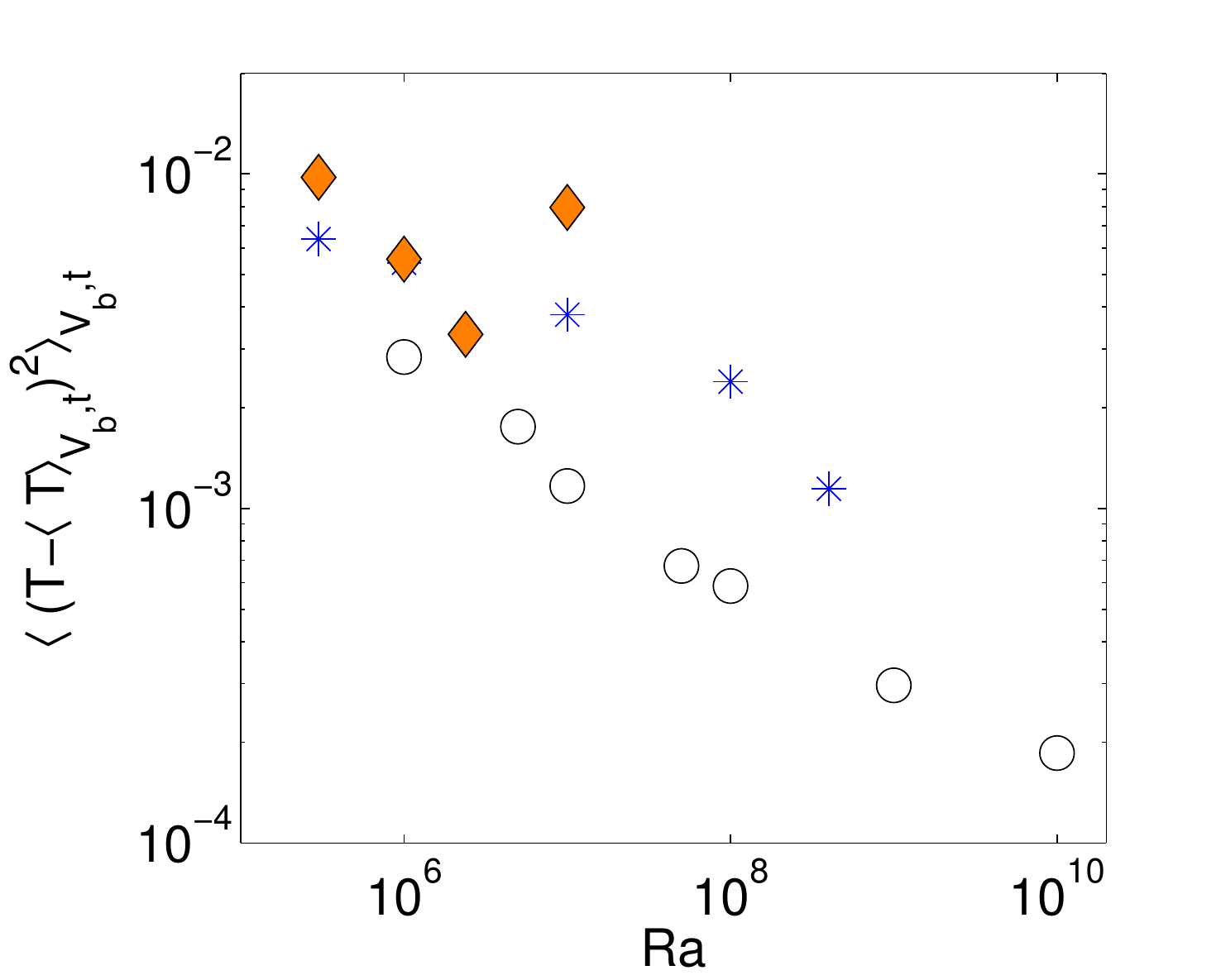}
\includegraphics[width=0.32\textwidth]{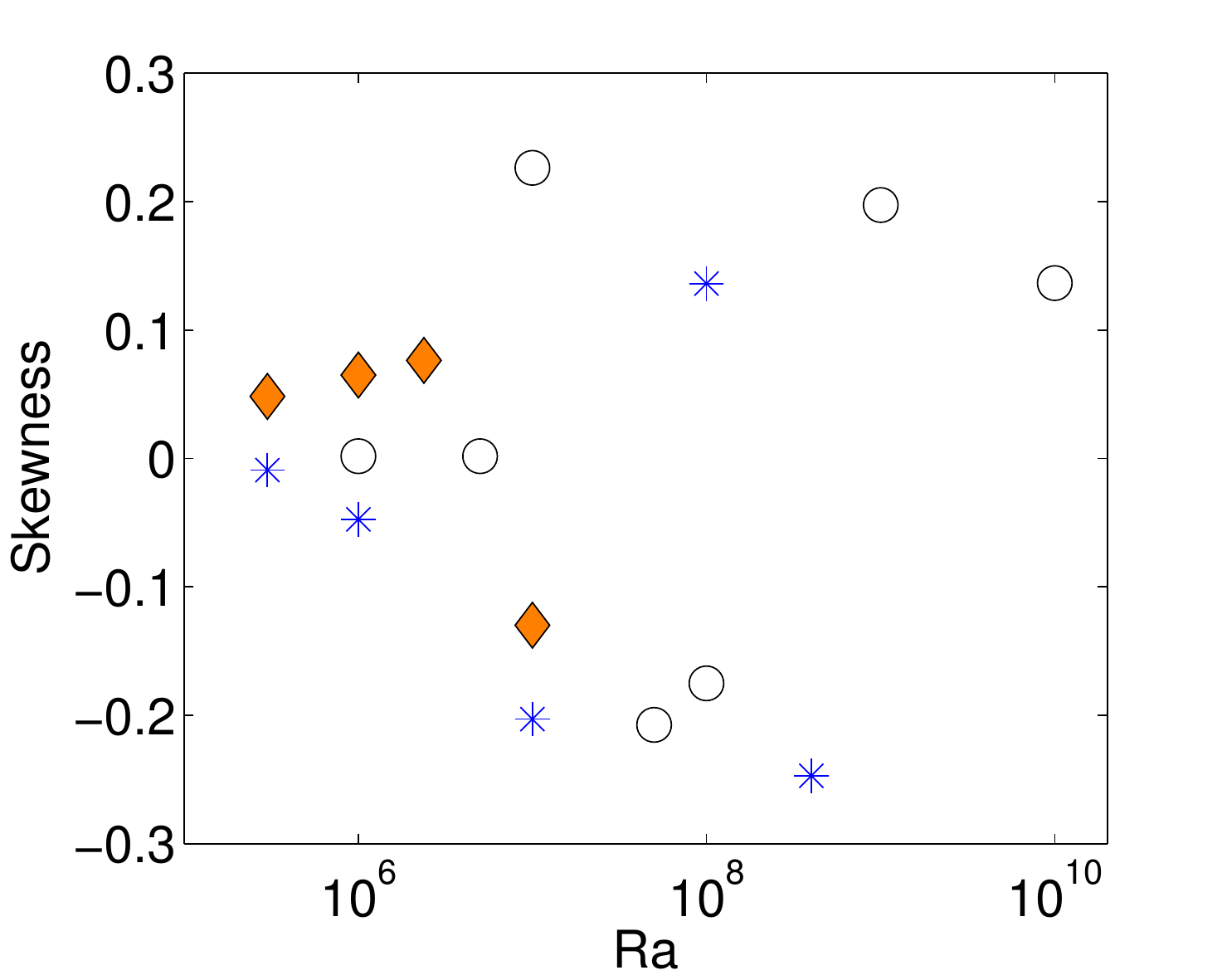}
\includegraphics[width=0.32\textwidth]{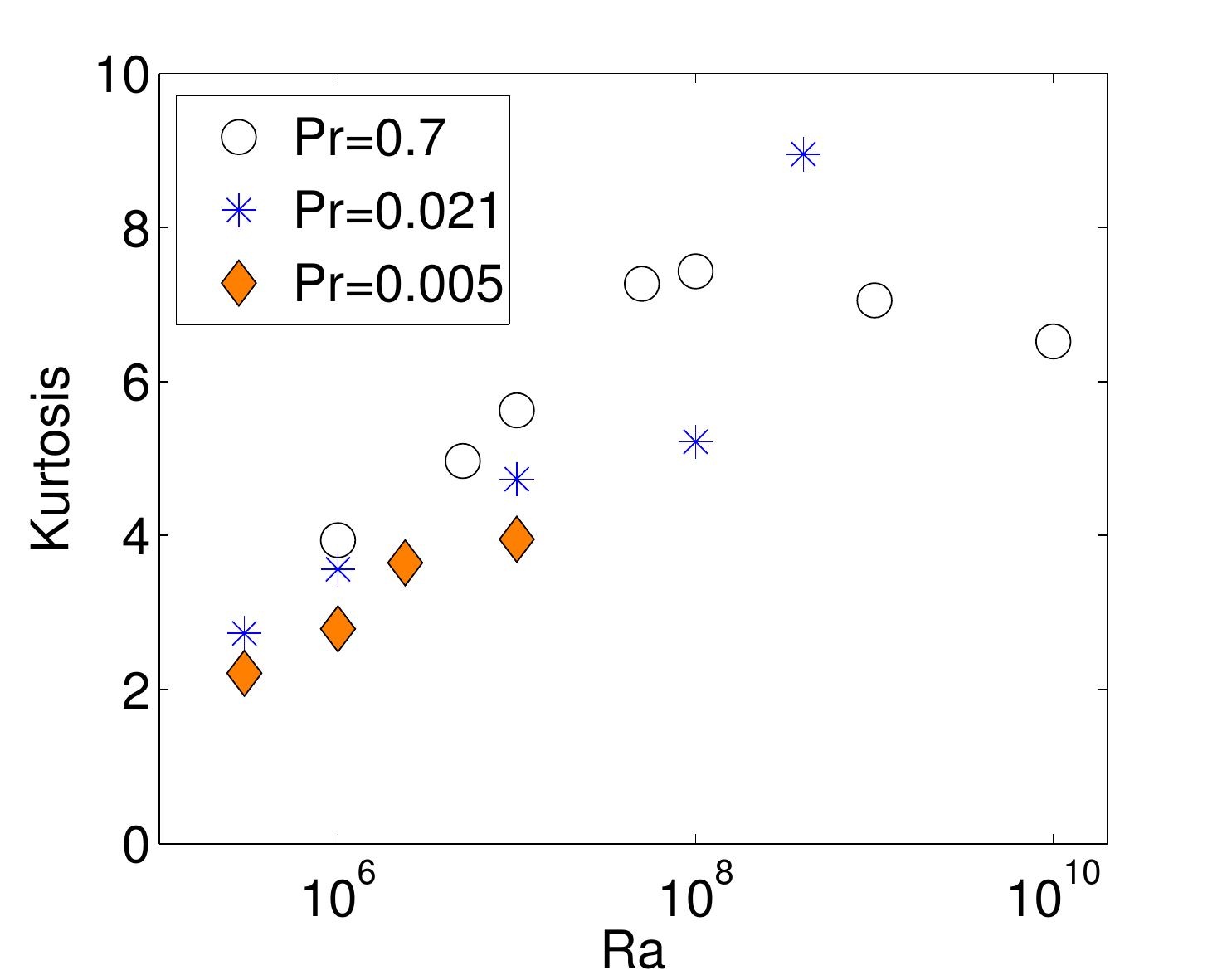}
\caption{Width $\langle (T-\langle T\rangle_{V_b,t})^2\rangle_{V_b,t}$, skewness and kurtosis calculated from the bulk temperature PDFs in the right column of Fig.~\ref{fign}. See Eqns. (\ref{skew}) and (\ref{kurt}) for our definitions of skewness and kurtosis.}
\label{fig18}
\end{figure}

The trend is confirmed in  Fig.~\ref{fig18}(a), where we plot the width for the bulk temperature PDFs. We see that the widths decrease systematically as 
Rayleigh increases except for run 19 which is an outlier. In Fig.~\ref{fig18}, we also plot the skewness and kurtosis for the bulk temperature pdfs as 
defined by the following equations:
\begin{eqnarray}
\label{skew}
      {\rm Skewness} & = &{ {\langle(T-\langle T\rangle_{V_b,t})^3\rangle_{V_b,t}}\over {\langle(T-\langle T\rangle_{V_b,t})^2\rangle^{3/2}_{V_b,t}}}\,,\\
        \label{kurt}
        {\rm Kurtosis} & = & {{\langle(T-\langle T\rangle_{V_b,t}\rangle_{V_b,t}^4}\over {\langle(T-\langle T\rangle_{V_b,t})^2\rangle_{V_b,t}^{2} }}\,.
\end{eqnarray}
The  skewness, or asymmetry of the PDF is fairly randomly scattered about zero, and the larger values for the higher Rayleigh numbers are most likely a result  of the shorter simulation times. The kurtosis, on the other hand, tends to grow systematically from sub-Gaussian values (smaller than 3) at the lowest Rayleigh 
numbers to values larger than 3 for all Prandtl numbers and sufficiently large $Ra$. This indicates a increasing intermittency of temperature statistics. The kurtosis
is found to level off for the series at $Pr=0.7$ for $Ra\gtrsim 10^8$.

\section{Summary and discussion}
The main purpose of the present work was summarize our DNS results for all Rayleigh and Prandtl numbers and then to use these results 
to predict a possible transition to turbulence in the viscous boundary layer of a turbulent
Rayleigh-B\'{e}nard convection flow by different analysis methods. We used our existing and further extended direct numerical simulation 
data basis for three-dimensional turbulent convection in a closed cylindrical cell at an aspect ratio of 1. Most of the data are obtained in the 
regime of very low Prandtl numbers, i.e., in a regime of convection where the thermal and viscous boundary layers are partly decoupled from 
each other. It is the thin viscous boundary layer which shows a transitional behavior that is accompanied by intermittent fluctuations of local 
thickness, wall stress or velocity amplitudes. In the focus of the present work was the analysis of ``standard'' quantities of turbulent boundary
layer analysis, such as the existence of logarithmic profiles, displacement and momentum displacement thicknesses and the resulting shape factors. 
Based on such quantities, we suggest three methods that predict a range of critical Rayleigh numbers for a transition to boundary layer turbulence
by extrapolation. Furthermore, a fourth method which we discussed in \cite{Scheel2016} is added for completeness.

One point that did not receive much attention in the past is related to the specific shape of the velocity profile. Theoretical models such 
as \cite{Grossmann2011} assume a logarithmic profile that extends all the way to the mid plane of the convection layer or cell. The mean streamwise
velocity profiles in our simulations are however found to be closer (although not equal) to a planar wall-jet with a pronounced local mean velocity maximum 
and an offset (see fig. \ref{fig7}). Similarly to channels, this class of flows obeys a logarithmic law of the wall as reported in refs. \citep{Eriksson1998,Ahlman2007}. It would thus be interesting to verify whether the decreasing mean velocity towards the bulk has consequences for the scaling predictions.
Which theoretical model best describes the viscous BL profiles near the plates and how this effects the global transport across the boundary layers 
is still an open question that  requires closer inspection.

The predicted transition ranges in all methods are shifted to lower Rayleigh number values as the Prandtl number decreases which is 
consistent with the results of the scaling theory by Grossmann and Lohse (see e.g. Fig. 3(a) in ref.\cite{Ahlers2009}). For convection in 
air, the predictions of methods I and IV are consistent with experimental results \cite{Ahlers2017}. The four methods that are independent of each 
other predict partly overlapping ranges of critical Rayleigh numbers. We note again that the extrapolations in all cases have large error bars. For 
the smallest Prandtl number, namely run 19, longer time series would be needed to resolve the slow transformation of the large-scale 
flow in the cell. The intention of this work is to suggest several analysis tools that can be applied in future simulations and partly even in experiments.
At this stage, it is hard to say which of our methods is most reliable, although it is seen that method III has the largest  uncertainty ranges
and method II provides only a lower bound as it predicts just when the boundary layer reaches the log-law.

We also wish to note that our predictions do 
not imply that such a transition must occur. 
Our study should therefore be understood as an encouragement to conduct a new series of high-Rayleigh number laboratory experiments, in particular in the low Prandtl 
number regime. This is to our view one of the most interesting sectors of the $Ra-Pr$ parameter plane, since the turbulence is highly inertial as we 
discussed here and
in our subsequent works \cite{Schumacher2015,Scheel2016,Schumacher2016}. Specifically, liquid sodium, the lowest Prandtl number that can be obtained in
a laboratory, seems to occupy a special role. It is still not well-understood if the turbulent convection flow switches into a different regime as Rayleigh number increases, as a result of the very diffusive temperature field. Our trends for the Nusselt number in Fig. \ref{fig3} might suggest such a behavior. Further longer-term 
numerical simulations are currently underway to understand these observations better and will be reported elsewhere.

\acknowledgements 
JDS acknowledges support by the Gerhard Mercator Fellowship of the Deutsche For\-schungs\-gemeinschaft within the Priority Programme
Turbulent Superstructures under Grant No. SPP 1881. We acknowledge supercomputing time at the Blue  Gene/Q JUQUEEN at the J\"ulich 
Supercomputing Centre which was provided by the large scale project grant HIL12 of the John von Neumann Institute for Computing. Furthermore, we acknowledge 
an award of computer time provided by the INCITE program. This research also used resources of the Argonne Leadership 
Computing Facility (ALCF) at Argonne National Laboratory, which is supported by the US Department of Energy
under contract DE-AC02-06CH11357. We also thank Guenter Ahlers, Christian Cierpka, Katepalli R. Sreenivasan, and Jared P. Whitehead
for helpful comments and discussions.

\section{Appendix: Transformation into LSC frame}
We summarize the rotations of coordinates and fields that have been used in Ref. \cite{Schumacher2016} to determine a streamwise and
spanwise in a closed cylindrical convection cell. The mean orientation angle of the LSC is calculated in each plane at fixed height $x_3>0$ by 
\begin{equation}
\label{angle}
\langle {\phi}(x_3,t)\rangle_b = \arctan \left [\frac{\langle u_2(x_3,t)\rangle_{b}}{\langle u_1(x_3,t)\rangle_b} \right]\,.
\end{equation}
At each plane of constant $x_3$ we apply afterwards 
\begin{equation}
  \label{rot}
  \left(\begin{array}{c}
                                \hat{x}_1 \\
                                \hat{x}_2 \\
                                \hat{x}_3                                
                                \end{array}\right) = 
   \left(\begin{array}{ccc}
                                \cos \langle \phi\rangle_b  & \sin \langle \phi\rangle_b  & 0 \\
                                -\sin \langle \phi\rangle_b  & \cos \langle \phi\rangle_b  & 0 \\
                                0 & 0 & 1
                                \end{array}\right) 
  \left(\begin{array}{c}
                                x_1 \\
                                x_2 \\
                                x_3                               
                                \end{array}\right)\;.               
\end{equation}
and thus the velocity components transform correspondingly as 
\begin{equation}
  \label{rot1}
  \left(\begin{array}{c}
                                \hat{u}_1 \\
                                \hat{u}_2 \\
                                \hat{u}_3                                
                                \end{array}\right) = 
   \left(\begin{array}{ccc}
                                \cos \langle \phi\rangle_b  & \sin \langle \phi\rangle_b  & 0 \\
                                -\sin \langle \phi\rangle_b  & \cos \langle \phi\rangle_b  & 0 \\
                                0 & 0 & 1
                                \end{array}\right) 
  \left(\begin{array}{c}
                                u_1 \\
                                u_2 \\
                                u_3                               
                                \end{array}\right)\;.               
\end{equation}

\end{document}